\documentclass[12pt,twoside, a4paper]{article}
\def\pd{\partial}
\def\mc{\mathcal}

\usepackage[dvips]{graphicx}
\usepackage{amssymb}
\usepackage{amssymb,amsmath}
\usepackage{graphicx}
\usepackage{dsfont}
\usepackage{caption}
\usepackage{subcaption}
\input{epsf.sty}  
\begin{document}
\begin{center}
\LARGE{\textbf{Supersymmetric $AdS_5$ black holes and strings from 5D $N=4$ gauged supergravity}}
\end{center}
\vspace{1 cm}
\begin{center}
\large{\textbf{H. L. Dao}$^a$ and \textbf{Parinya Karndumri}$^b$}
\end{center}
\begin{center}
$^a$Department of Physics, National University of Singapore,
3 Science Drive 2, Singapore 117551\\
E-mail: hl.dao@u.nus.edu\\
$^b$String Theory and Supergravity Group, Department
of Physics, Faculty of Science, Chulalongkorn University, 254 Phayathai Road, Pathumwan, Bangkok 10330, Thailand \\
E-mail: parinya.ka@hotmail.com \vspace{1 cm}
\end{center}
\begin{abstract}
We study supersymmetric $AdS_3\times \Sigma_2$ and $AdS_2\times \Sigma_3$ solutions, with $\Sigma_2=S^2,H^2$ and $\Sigma_3=S^3,H^3$, in five-dimensional $N=4$ gauged supergravity coupled to five vector multiplets. The gauge groups considered here are $U(1)\times SU(2)\times SU(2)$, $U(1)\times SO(3,1)$ and $U(1)\times SL(3,\mathbb{R})$. For $U(1)\times SU(2)\times SU(2)$ gauge group admitting two supersymmetric $N=4$ $AdS_5$ vacua, we identify a new class of $AdS_3\times \Sigma_2$ and $AdS_2\times H^3$ solutions preserving four supercharges. Holographic RG flows describing twisted compactifications of $N=2$ four-dimensional SCFTs dual to the $AdS_5$ vacua to the SCFTs in two and one dimensions dual to these geometries are numerically given. The solutions can also be interpreted as supersymmetric black strings and black holes in asymptotically $AdS_5$ spaces with near horizon geometries given by $AdS_3\times \Sigma_2$ and $AdS_2\times H^3$, respectively. These solutions broaden previously known black brane solutions including half-supersymmetric $AdS_5$ black strings recently found in $N=4$ gauged supergravity. Similar solutions are also studied in non-compact gauge groups $U(1)\times SO(3,1)$ and $U(1)\times SL(3,\mathbb{R})$.      
\end{abstract}
\newpage
\section{Introduction}
Black branes of different spatial dimensions play an important role in the develoment of string/M-theory. They lead to many insightful results such as the construction of gauge theories in various dimensions and the celebrated AdS/CFT correspondence \cite{maldacena}. According to the latter, black branes in asymptotically $AdS$ spaces are of particular interest since they are dual to RG flows across dimensions from superconformal field theories (SCFTs) dual to the asymptotically $AdS$ spaces to lower-dimensional fixed points dual to the near horizon geometries \cite{Maldacena_Nunez_nogo}. Recently, a new approach for computing microscopic entropy of $AdS_4$ balck holes has been introduced based on twisted partition functions of three-dimensional SCFTs \cite{BH_entropy_Zaffaroni1,BH_entropy_Zaffaroni1_2,BH_entropy_Zaffaroni2,BH_entropy_Zaffaroni2_1,BH_entropy_Zaffaroni3,BH_entropy_IIA1,
BH_entropy_IIA2,Bobev_AdS4_BH,BH_microstate_Cabo}. This has also been applied to $AdS$ black holes in other dimensions \cite{BH_microstate_7D,BH_microstate_6D,BH_micro_5D_ Zaffaroni,Minwoo_6D_BH2,BH_microstate_6D2,BH_microstate_5D1,5D_BH_microstate_Cabo}.       
\\
\indent In this paper, we are interested in supersymmetric black holes and black strings in asymptocally $AdS_5$ spaces from five-dimensional $N=4$ gauged supergravity coupled to vector multiplets constructed in \cite{N4_gauged_SUGRA,5D_N4_Dallagata} using the embedding tensor formalism \cite{maximal_gauged_SUGRA,maximal_5Dgauged_SUGRA,maximal_3Dgauged_SUGRA}. These solutions have near horizon geometries of the forms $AdS_2\times \Sigma_3$ and $AdS_3\times \Sigma_2$, respectively. We will consider $\Sigma_3$ in the form of a three-sphere ($S^3$) and a three-dimensional hyperbolic space ($H^3$). Similarly, $\Sigma_2$ will be given by a two-sphere ($S^2$) and a two-dimensional hyperbolic space ($H^2$), or a Riemann surface of genus $\mathfrak{g}>1$. Similar solutions have previously been found in minimal and maximal gauged supergravities, see for example \cite{black_string_Klemm1,black_string_Klemm2,black_string_Klemm3,black_string_Klemm4,
black_string_Klemm5,5D_N2_BH,5D_N8_BH,5D_N8_BH_Minwoo,MMSJ_AdS3}. This type of solutions has also appeared in pure $N=4$ gauged supergravity in \cite{5D_N4_Romans}, and recently, half-supersymmetric black strings with hyperbolic horizons have been found in matter-coupled $N=4$ gauged supergravity with compact $U(1)\times SU(2)\times SU(2)$ and non-compact $U(1)\times SO(3,1)$ gauge groups \cite{5D_N4_flow}.   
\\
\indent We will look for more general solutions of $AdS_5$ black strings with both hyperbolic and spherical horizons and preserving $\frac{1}{4}$ of the $N=4$ supersymmetry in five dimensions. The solutions interpolate between $N=4$ supersymmetric $AdS_5$ vacua of the gauged supergravity and near horizon geometries of the form $AdS_3\times \Sigma_2$. In addition, we will look for supersymmetric black holes interpolating between $AdS_5$ vacua and near horizon geometries $AdS_2\times \Sigma_3$. According to the AdS/CFT correspondence, these solutions describe RG flows across dimensions from the dual $N=2$ SCFTs to two- and one-dimensional SCFTs in the IR. The IR SCFTs are obtained via twisted compactifications of $N=2$ SCFTs in four dimensions. Many solutions of this type have been found in various space-time dimensions, see \cite{Cucu_AdSD-2,BB,Wraped_D3,3D_CFT_from_LS_point,flow_acrossD_bobev,BBC,
4D_SCFT_from_M5,Wraped_M5,N3_AdS2,flow_across_Betti,AdS2_trisasakian,7D_twist,6D_twist} for an incomplete list.
\\
\indent We mainly consider $N=4$ gauged supergravity coupled to five vector multiplets with gauge groups entirely embedded in the global symmetry $SO(5,5)$. We will also restrict ourselves to gauge groups that lead to supersymmetric $AdS_5$ vacua. These gauge groups have been shown in \cite{AdS5_N4_Jan} to take the form of $U(1)\times H_0\times H$ with the $U(1)$ gauged by the graviphoton that is a singlet under $USp(4)\sim SO(5)$ R-symmetry. The $H\subset SO(n+3-\textrm{dim}\, H_0)$ is a compact group gauged by vector fields in the vector multiplets, and $H_0$ is a non-compact group gauged by three of the graviphotons and $\textrm{dim}\, H_0-3$ vectors from the vector multiplets. The remaining two graviphotons in the fundamental representation of $SO(5)$ are dualized to massive two-form fields. In addition, $H_0$ must contain an $SU(2)$ subgroup. For the case of five vector multiplets, possible gauge groups that admit supersymmetric $AdS_5$ vacua and can be embedded in $SO(5,5)$ are $U(1)\times SU(2)\times SU(2)$, $U(1)\times SO(3,1)$ and $U(1)\times SL(3,\mathbb{R})$. We will look for $AdS_5$ black string and black hole solutions in all of these gauge groups. 
\\
\indent The paper is organized as follow. In section \ref{N4_SUGRA},
we review $N=4$ gauged supergravity in five dimensions coupled to vector multiplets using the embedding tensor formalism. In section \ref{U1_SU2_SU2_gauge_group}, we find supersymmetric $AdS_3\times \Sigma_2$ solutions preserving four supercharges and give numerical RG flow solutions interpolating between these geometries and supersymmetric $AdS_5$ vacua. An $AdS_2\times H^3$ solution together with an RG flow interpolating between $AdS_5$ vacua and this geometry will also be given. In section \ref{U1_SO3_1_gauge_group} and \ref{U1_SL3_R_gauge_group}, we repeat the same analysis for non-compact $U(1)\times SO(3,1)$ and $U(1)\times SL(3,\mathbb{R})$ gauge groups. Since the $U(1)\times SL(3,\mathbb{R})$ gauge group has not been studied in \cite{5D_N4_flow}, we will discuss its construction and supersymmetric $AdS_5$ vacuum in detail. The full scalar mass spectrum at this critical point will also be given. This should be useful in the holographic context since it contains information on dimensions of operators dual to supergravity scalars. We end the paper with some conclusions and comments in section \ref{conclusion}.

\section{Five dimensional $N=4$ gauged supergravity coupled to vector multiplets}\label{N4_SUGRA} 
In this section, we briefly review the structure of five dimensional $N=4$ gauged supergravity coupled to vector multiplets with the emphasis on formulae relevant for finding supersymmetric solutions. The detailed construction of $N=4$ gauged supergravity can be found in \cite{N4_gauged_SUGRA} and \cite{5D_N4_Dallagata}. 
\\
\indent The $N=4$ gravity multiplet consists of the graviton
$e^{\hat{\mu}}_\mu$, four gravitini $\psi_{\mu i}$, six vectors $A^0$ and
$A_\mu^m$, four spin-$\frac{1}{2}$ fields $\chi_i$ and one real
scalar $\Sigma$, the dilaton. Space-time and tangent space indices are denoted respectively by $\mu,\nu,\ldots =0,1,2,3,4$ and
$\hat{\mu},\hat{\nu},\ldots=0,1,2,3,4$. The $SO(5)\sim USp(4)$
R-symmetry indices are described by $m,n=1,\ldots, 5$ for the
$SO(5)$ vector representation and $i,j=1,2,3,4$ for the $SO(5)$
spinor or $USp(4)$ fundamental representation. The gravity multiplet can couple to an arbitrary number $n$ of vector multiplets. Each vector multiplet contains a vector field $A_\mu$, four gaugini $\lambda_i$ and five scalars $\phi^m$. The $n$ vector
multiplets will be labeled by indices $a,b=1,\ldots, n$, and the components fields within these vector multiplets will be denoted by $(A^a_\mu,\lambda^{a}_i,\phi^{ma})$. From both gravity and vector multiplets, there are in total $6+n$ vector fields which will be denoted by $A^{\mc{M}}_\mu=(A^0_\mu,A^m_\mu,A^a_\mu)$. All fermionic fields are described by symplectic Majorana spinors subject to the following condition
\begin{equation}
\xi_i=\Omega_{ij}C(\bar{\xi}^j)^T 
\end{equation}
with $C$ and $\Omega_{ij}$ being respectively the charge conjugation matrix and $USp(4)$ symplectic form. 
\\
\indent The $5n$ scalar fields from the vector multiplets parametrize the $SO(5,n)/SO(5)\times SO(n)$ coset. To describe this coset manifold, we introduce a coset representative $\mc{V}_M^{\phantom{M}A}$ transforming under the global $SO(5,n)$ and the local $SO(5)\times SO(n)$ by left and right multiplications, respectively. We use indices $M,N,\ldots=1,2,\ldots , 5+n$ for global $SO(5,n)$ indices. The local $SO(5)\times SO(n)$ indices $A,B,\ldots$ will be split into $A=(m,a)$. We can accordingly write the coset representative as
\begin{equation}
\mc{V}_M^{\phantom{M}A}=(\mc{V}_M^{\phantom{M}m},\mc{V}_M^{\phantom{M}a}).
\end{equation}
The matrix $\mc{V}_M^{\phantom{M}A}$ is an element of $SO(5,n)$ and satisfies the relation
\begin{equation}
\eta_{MN}={\mc{V}_M}^A{\mc{V}_N}^B\eta_{AB}=-\mc{V}_M^{\phantom{M}m}\mc{V}_N^{\phantom{M}m}+\mc{V}_M^{\phantom{M}a}\mc{V}_N^{\phantom{M}a}
\end{equation}
with $\eta_{MN}=\textrm{diag}(-1,-1,-1,-1,-1,1,\ldots,1)$ being the $SO(5,n)$ invariant tensor. Equivalently, the $SO(5,n)/SO(5)\times SO(n)$ coset can also be described in term of a symmetric matrix
\begin{equation}
M_{MN}=\mc{V}_M^{\phantom{M}m}\mc{V}_N^{\phantom{M}m}+\mc{V}_M^{\phantom{M}a}\mc{V}_N^{\phantom{M}a}
\end{equation}
which is manifestly invariant under the $SO(5)\times SO(n)$ local symmetry.
\\
\indent Gaugings promote a given subgroup $G_0$ of the full global symmetry $SO(1,1)\times SO(5,n)$ of $N=4$ supergravity coupled to $n$ vector multiplets to be a local symmetry. These gaugings are efficiently described by using the embedding tensor formalism. $N=4$ supersymmetry allows three components of the embedding tensor $\xi^{M}$, $\xi^{MN}=\xi^{[MN]}$ and $f_{MNP}=f_{[MNP]}$ \cite{N4_gauged_SUGRA}. The first component $\xi^M$ describes the embedding of the gauge group in the $SO(1,1)\sim \mathbb{R}^+$ factor identified with the coset space parametrized by the dilaton $\Sigma$. From the result of \cite{AdS5_N4_Jan}, the existence of $N=4$ supersymmetric $AdS_5$ vacua requires $\xi^M=0$. In this paper, we are only interested in solutions that are asymptotically $AdS_5$, so we will restrict ourselves to the gaugings with $\xi^{M}=0$.
\\
\indent For $\xi^{M}=0$, the gauge group is entirely embedded in $SO(5,n)$ with the gauge generators given by 
\begin{equation}
{(X_M)_N}^P=-{f_M}^{QR}{(t_{QR})_N}^P={f_{MN}}^P\quad \textrm{and}\quad {(X_0)_N}^P=-\xi^{QR}{(t_{QR})_N}^P={\xi_N}^P\, .
\end{equation}
The matrices ${(t_{MN})_P}^Q=\delta^Q_{[M}\eta_{N]P}$ are $SO(5,n)$ generators in the fundamental representation. The full covariant derivative reads
\begin{equation}
D_\mu=\nabla_\mu+A_\mu^{M}X_M+A^0_\mu X_0
\end{equation}
where $\nabla_\mu$ is the usual space-time covariant derivative. We use the convention that the definition of $\xi^{MN}$ and $f_{MNP}$ includes the gauge coupling constants. Note also that $SO(5,n)$ indices $M,N,\ldots$ are lowered and raised by $\eta_{MN}$ and its inverse $\eta^{MN}$, respectively. 
\\
\indent Generators $X_{\mc{M}}=(X_0,X_M)$ of a consistent gauge group must form a closed subalgebra of $SO(5,n)$. This requires $\xi^{MN}$ and $f_{MNP}$ to satisfy the quadratic constraints, see \cite{N4_gauged_SUGRA},
\begin{equation}
f_{R[MN}{f_{PQ]}}^R=0\qquad \textrm{and}\qquad {\xi_M}^Qf_{QNP}=0\, .
\end{equation}      
Gauge groups that admit $N=4$ supersymmetric $AdS_5$ vacua generally take the form of $U(1)\times H_0\times H$, see \cite{AdS5_N4_Jan} for more detail. The $U(1)$ is gauged by $A^0_\mu$ while $H\subset SO(n+3-\textrm{dim}\, H_0)$ is a compact group gauged by vector fields in the vector multiplets. $H_0$ is a non-compact group gauged by three of the graviphotons and $\textrm{dim}\, H_0-3$ vectors from the vector multiplets. $H_0$ must also contain an $SU(2)$ subgroup. For simple groups, $H_0$ can be $SU(2)\sim SO(3)$, $SO(3,1)$ and $SL(3,\mathbb{R})$. 
\\
\indent In the embedding tensor formalism, there are two-form fields $B_{\mu\nu \mc{M}}$ that are introduced off-shell. These two-form fields do not have kinetic terms and couple to vector fields via a topological term. They satisfy a first-order field equation given by, see \cite{N4_gauged_SUGRA} for more detail,
\begin{equation}
\xi^{\mc{M}\mc{N}}\left[\frac{1}{6\sqrt{2}}\epsilon_{\mu\nu\rho\lambda\sigma}\mc{H}^{(3)\rho\lambda\sigma}_{\mc{N}}-\mc{M}_{\mc{N}\mc{P}}
\mc{H}^{\mc{P}}_{\mu\nu}\right]=0\label{2-form_field_eq}
\end{equation}
in which $\mc{M}_{00}=\Sigma^{-4}$, $\mc{M}_{0M}=0$ and $\mc{M}_{MN}=\Sigma^2M_{MN}$. The field strength $\mc{H}^{(3)}_{\mc{M}}$ is defined by
\begin{equation}
\xi^{\mc{M}\mc{N}}\mc{H}^{(3)}_{\mu\nu\rho\mc{N}}=\xi^{\mc{M}\mc{N}}\left[3D_{[\mu}B_{\nu\rho]\mc{N}}
+6d_{\mc{NPQ}}A^{\mc{P}}_{[\mu}\left(\pd_\nu A^{\mc{Q}}_{\rho]}+\frac{1}{3}{X_{\mc{RS}}}^{\mc{Q}}A^{\mc{R}}_\nu A^{\mc{S}}_{\rho]}\right)\right]\label{H3_def}
\end{equation}
with $d_{0MN}=d_{MN0}=d_{M0N}=\eta_{MN}$ and 
\begin{equation}
{X_{MN}}^P=-{f_{MN}}^P,\qquad {X_{M0}}^0=0,\qquad {X_{0M}}^N=-{\xi_M}^N\, . 
\end{equation}
In all of the solutions considered here, the Chern-Simons term in equation \eqref{H3_def} vanish due to a particular form of the ansatz for the gauge fields. In addition, the term $\mc{M}_{\mc{N}\mc{P}}\mc{H}^{\mc{P}}_{\mu\nu}$ in equation \eqref{2-form_field_eq} also vanish provided that the gauge fields $A^1$ and $A^2$ are set to zero. With all these, the two-form fields can be consistently truncated out. We will accordingly set all the two-form fields to zero from now on. 
\\
\indent The bosonic Lagrangian of a general gauged $N=4$ supergravity coupled to $n$ vector multiplets can accordingly be written as
\begin{eqnarray}
e^{-1}\mc{L}&=&\frac{1}{2}R-\frac{1}{4}\Sigma^2M_{MN}\mc{H}^M_{\mu\nu}\mc{H}^{N\mu\nu}-\frac{1}{4}\Sigma^{-4}\mc{H}^0_{\mu\nu}\mc{H}^{0\mu\nu}\nonumber \\
& &-\frac{3}{2}\Sigma^{-2}D_\mu \Sigma D^\mu \Sigma +\frac{1}{16} D_\mu M_{MN}D^\mu
M^{MN}-V+e^{-1}\mc{L}_{\textrm{top}}
\end{eqnarray}
where $e$ is the vielbein determinant. $\mc{L}_{\textrm{top}}$ is the topological term whose explicit form will not be given here since, given our ansatz for the gauge fields, it will not play any role in the present discussion. 
\\
\indent With vanishing two-form fields, the covariant gauge field strength tensors read
\begin{equation}
\mc{H}^{\mc{M}}_{\mu\nu}=2\pd_{[\mu}A^{\mc{M}}_{\nu]}+{X_{\mc{N}\mc{P}}}^{\mc{M}}A^{\mc{N}}_\mu A^{\mc{P}}_\nu\, .\label{covariant_field_strength}
\end{equation}
The scalar potential is given by
\begin{eqnarray}
V&=&-\frac{1}{4}\left[f_{MNP}f_{QRS}\Sigma^{-2}\left(\frac{1}{12}M^{MQ}M^{NR}M^{PS}-\frac{1}{4}M^{MQ}\eta^{NR}\eta^{PS}\right.\right.\nonumber \\
& &\left.+\frac{1}{6}\eta^{MQ}\eta^{NR}\eta^{PS}\right) +\frac{1}{4}\xi_{MN}\xi_{PQ}\Sigma^4(M^{MP}M^{NQ}-\eta^{MP}\eta^{NQ})\nonumber \\
& &\left.
+\frac{\sqrt{2}}{3}f_{MNP}\xi_{QR}\Sigma M^{MNPQRS}\right]
\end{eqnarray}
where $M^{MN}$ is the inverse of $M_{MN}$, and $M^{MNPQRS}$ is obtained from
\begin{equation}
M_{MNPQR}=\epsilon_{mnpqr}\mc{V}_{M}^{\phantom{M}m}\mc{V}_{N}^{\phantom{M}n}
\mc{V}_{P}^{\phantom{M}p}\mc{V}_{Q}^{\phantom{M}q}\mc{V}_{R}^{\phantom{M}r}
\end{equation}
by raising the indices with $\eta^{MN}$. 
\\
\indent Supersymmetry transformations of fermionic fields $(\psi_{\mu i},\chi_i,\lambda^a_i)$ are given by
\begin{eqnarray}
\delta\psi_{\mu i} &=&D_\mu \epsilon_i+\frac{i}{\sqrt{6}}\Omega_{ij}A^{jk}_1\gamma_\mu\epsilon_k\nonumber \\
& &-\frac{i}{6}\left(\Omega_{ij}\Sigma{\mc{V}_M}^{jk}\mc{H}^M_{\nu\rho}-\frac{\sqrt{2}}{4}\delta^k_i\Sigma^{-2}\mc{H}^0_{\nu\rho}\right)({\gamma_\mu}^{\nu\rho}-4\delta^\nu_\mu\gamma^\rho)\epsilon_k,\\
\delta \chi_i &=&-\frac{\sqrt{3}}{2}i\Sigma^{-1} D_\mu
\Sigma\gamma^\mu \epsilon_i+\sqrt{2}A_2^{kj}\epsilon_k\nonumber \\
& &-\frac{1}{2\sqrt{3}}\left(\Sigma \Omega_{ij}{\mc{V}_M}^{jk}\mc{H}^M_{\mu\nu}+\frac{1}{\sqrt{2}}\Sigma^{-2}\delta^k_i\mc{H}^0_{\mu\nu}\right)\gamma^{\mu\nu}\epsilon_k,\\
\delta \lambda^a_i&=&i\Omega^{jk}({\mc{V}_M}^aD_\mu
{\mc{V}_{ij}}^M)\gamma^\mu\epsilon_k+\sqrt{2}\Omega_{ij}A_{2}^{akj}\epsilon_k-\frac{1}{4}\Sigma{\mc{V}_M}^a\mc{H}^M_{\mu\nu}\gamma^{\mu\nu}\epsilon_i
\end{eqnarray}
in which the fermion shift matrices are defined by
\begin{eqnarray}
A_1^{ij}&=&-\frac{1}{\sqrt{6}}\left(\sqrt{2}\Sigma^2\Omega_{kl}{\mc{V}_M}^{ik}{\mc{V}_N}^{jl}\xi^{MN}+\frac{4}{3}\Sigma^{-1}{\mc{V}^{ik}}_M{\mc{V}^{jl}}_N{\mc{V}^P}_{kl}{f^{MN}}_P\right),\nonumber
\\
A_2^{ij}&=&\frac{1}{\sqrt{6}}\left(\sqrt{2}\Sigma^2\Omega_{kl}{\mc{V}_M}^{ik}{\mc{V}_N}^{jl}\xi^{MN}-\frac{2}{3}\Sigma^{-1}{\mc{V}^{ik}}_M{\mc{V}^{jl}}_N{\mc{V}^P}_{kl}{f^{MN}}_P\right),\nonumber
\\
A_2^{aij}&=&-\frac{1}{2}\left(\Sigma^2{\mc{V}_M}^a{\mc{V}_N}^{ij}\xi^{MN}-\sqrt{2}\Sigma^{-1}\Omega_{kl}{\mc{V}_M}^a{\mc{V}_N}^{ik}{\mc{V}_P}^{jl}f^{MNP}\right).
\end{eqnarray}
\indent In these equations, $\mc{V}_M^{\phantom{M}ij}$ is defined in term of ${\mc{V}_M}^m$ as
\begin{equation}
{\mc{V}_M}^{ij}=\frac{1}{2}{\mc{V}_M}^{m}\Gamma^{ij}_m
\end{equation}
where $\Gamma^{ij}_m=\Omega^{ik}{\Gamma_{mk}}^j$ and ${\Gamma_{mi}}^j$ are $SO(5)$ gamma matrices. Similarly, the inverse element ${\mc{V}_{ij}}^M$ can be written as
\begin{equation}
{\mc{V}_{ij}}^M=\frac{1}{2}{\mc{V}_m}^M(\Gamma^{ij}_m)^*=\frac{1}{2}{\mc{V}_m}^M\Gamma_{m}^{kl}\Omega_{ki}\Omega_{lj}\,
.
\end{equation}
In the subsequent analysis, we use the following explicit choice of $SO(5)$ gamma matrices ${\Gamma_{mi}}^j$ given by
\begin{eqnarray}
\Gamma_1&=&-\sigma_2\otimes \sigma_2,\qquad \Gamma_2=i\mathbb{I}_2\otimes \sigma_1,\qquad \Gamma_3=\mathbb{I}_2\otimes \sigma_3,\nonumber\\
\Gamma_4&=&\sigma_1\otimes \sigma_2,\qquad \Gamma_5=\sigma_3\otimes \sigma_2
\end{eqnarray}
where $\sigma_i$, $i=1,2,3$ are the usual Pauli matrices.
\\
\indent The covariant derivative on $\epsilon_i$ reads
\begin{equation}
D_\mu \epsilon_i=\pd_\mu \epsilon_i+\frac{1}{4}\omega_\mu^{ab}\gamma_{ab}\epsilon_i+{Q_{\mu i}}^j\epsilon_j
\end{equation}
where the composite connection is defined by
\begin{equation}
{Q_{\mu i}}^j={\mc{V}_{ik}}^M\pd_\mu {\mc{V}_M}^{kj}-A^0_\mu\xi^{MN}\mc{V}_{Mik}{\mc{V}_N}^{kj}-A^M_\mu{\mc{V}_{ik}}^N\mc{V}^{kjP}f_{MNP}\, .
\end{equation}
\indent In this work, we mainly focus on the case of $n=5$ vector multiplets. To parametrize the scalar coset $SO(5,5)/SO(5)\times SO(5)$, it is useful to introduce a basis for $GL(10,\mathbb{R})$ matrices
\begin{equation} 
(e_{MN})_{PQ}=\delta_{MP}\delta_{NQ}
\end{equation}
in terms of which $SO(5,5)$ non-compact generators are given by
\begin{equation}
Y_{ma}=e_{m,a+5}+e_{a+5,m},\qquad m=1,2,\ldots, 5,\qquad a=1,2,\ldots, 5\, .
\end{equation}

\section{$U(1)\times SU(2)\times SU(2)$ gauge group}\label{U1_SU2_SU2_gauge_group}
For a compact $U(1)\times SU(2)\times SU(2)$ gauge group, components of the embedding tensor are given by
\begin{eqnarray}
\xi^{MN}&=&g_1(\delta^M_2\delta^N_1-\delta^M_1\delta^N_2),\\ 
f_{\tilde{m}+2,\tilde{n}+2,\tilde{p}+2}&=&-g_2\epsilon_{\tilde{m}\tilde{n}\tilde{p}},\qquad \tilde{m},\tilde{n},\tilde{p}=1,2,3,\\
f_{abc}&=&g_3\epsilon_{abc},\qquad a,b,c=1,2,3
\end{eqnarray} 
where $g_1$, $g_2$ and $g_3$ are the coupling constants for each factor in $U(1)\times SU(2)\times SU(2)$.
\\
\indent The scalar potential obtained from truncating the scalars from vector multiplets to $U(1)\times SU(2)_{\textrm{diag}}\subset U(1)\times SU(2)\times SU(2)$ singlets has been studied in \cite{5D_N4_flow}. There is one $U(1)\times SU(2)_{\textrm{diag}}$ singlet from the $SO(5,5)/SO(5)\times SO(5)$ coset corresponding to the following $SO(5,5)$ non-compact generator
\begin{equation}
Y_s=Y_{31}+Y_{42}+Y_{53}\, .
\end{equation}
With the coset representative given by 
\begin{equation}
\mc{V}=e^{\phi Y_s},\label{U1_SU2d_coset}
\end{equation}
the scalar potential can be computed to be
\begin{eqnarray}
V&=&\frac{1}{32\Sigma^2}\left[32\sqrt{2}g_1g_2\Sigma^3\cosh^3\phi-9(g_2^2+g_3^2)\cosh(2\phi) \right.\nonumber \\
& &-8(g_2^2-g_3^2-4\sqrt{2}g_1g_3\Sigma^3\sinh^3\phi-g_2g_3\sinh^3\phi)\nonumber \\ 
& &\left.+(g_2^2+g_3^2)\cosh(6\phi)\right].\label{potential_SU2d}
\end{eqnarray}
\\
\indent The potential admits two $N=4$ supersymmetric $AdS_5$ critical points given by
\begin{eqnarray}
\textrm{i}&:& \qquad \phi=0,\qquad \Sigma=1,\qquad V_0=-3g_1^2 \label{AdS5_1_compact}\\
\textrm{ii}&:&\qquad \phi=\frac{1}{2}\ln\left[\frac{g_3-g_2}{g_3+g_2}\right],\qquad \Sigma=\left(\frac{g_2g_3}{g_1\sqrt{2(g_3^2-g_2^2)}}\right)^{\frac{1}{3}},\nonumber \\ 
& &\qquad V_0=-3\left(\frac{g_1g_2^2g_3^2}{2(g_3^2-g_2^2)}\right)^{\frac{2}{3}}\, .\label{AdS5_2_compact}
\end{eqnarray}
In critical point i, we have set $g_2=-\sqrt{2}g_1$ to make this critical point occur at $\Sigma=1$. However, we will keep $g_2$ explicit in most expressions for brevity. Critical point i is invariant under the full gauge symmetry $U(1)\times SU(2)\times SU(2)$ while critical point ii preserves only $U(1)\times SU(2)_{\textrm{diag}}$ symmetry due to the non-vanising scalar $\phi$. $V_0$ denotes the cosmological constant, the value of the scalar potential at a critical point. 

\subsection{Supersymmetric black strings}
We now consider vacuum solutions of the form $AdS_3\times \Sigma_2$ with $\Sigma_2$ being $S^2$ or $H^2$. A number of $AdS_3\times H^2$ solutions that preserve eight supercharges together with RG flows interpolating between them and supersymmetric $AdS_5$ critical points have already been given in \cite{5D_N4_flow}. In this section, we look for more general solutions that preserve only four supercharges.  
\\
\indent We begin with the metric ansatz for the $\Sigma_2=S^2$ case
\begin{equation}
ds^2=e^{2f(r)}dx^2_{1,1}+dr^2+e^{2g(r)}(d\theta^2+\sin^2\theta d\phi^2)\label{metric_ansatz_AdS3S2}
\end{equation}
where $dx^2_{1,1}$ is the flat metric in two dimensions. For $\Sigma_2=H^2$, the metric is given by
\begin{equation}
ds^2=e^{2f(r)}dx^2_{1,1}+dr^2+e^{2g(r)}(d\theta^2+\sinh^2\theta d\phi^2).\label{metric_ansatz_AdS3H2}
\end{equation}
As $r\rightarrow \infty$, the metric becomes locally $AdS_5$ with $f(r)\sim g(r)\sim \frac{r}{L_{AdS_5}}$ while the near horizon geometry is characterized by the conditions $f(r)\sim \frac{r}{L_{AdS_3}}$ and constant $g(r)$, or equivalently $g'(r)=0$.  
\\
\indent To preserve some amount of supersymmetry, we perform a twist by cancelling the spin connection along the $\Sigma_2$ by some suitable choice of gauge fields. We will first consider abelian twists from the $U(1)\times U(1)\times U(1)$ subgroup of the $U(1)\times SU(2)\times SU(2)$ gauge symmetry. The gauge fields corresponding to this subgroup will be denoted by $(A^0,A^5,A^8)$. The ansatz for these gauge fields will be chosen as
\begin{equation}
A^{\mc{M}=0,5,8}=a_{\mc{M}}\cos\theta d\phi\, .\label{vector_ansatz_S2}
\end{equation} 
for the $S^2$ case and 
\begin{equation}
A^{\mc{M}=0,5,8}=a_{\mc{M}}\cosh\theta d\phi\, .\label{vector_ansatz_H2}
\end{equation}
for the $H^2$ case. 

\subsubsection{Solutions with $U(1)\times U(1)\times U(1)$ symmetry}  
\begin{figure}
         \centering
         \begin{subfigure}[b]{0.3\textwidth}
                 \includegraphics[width=\textwidth]{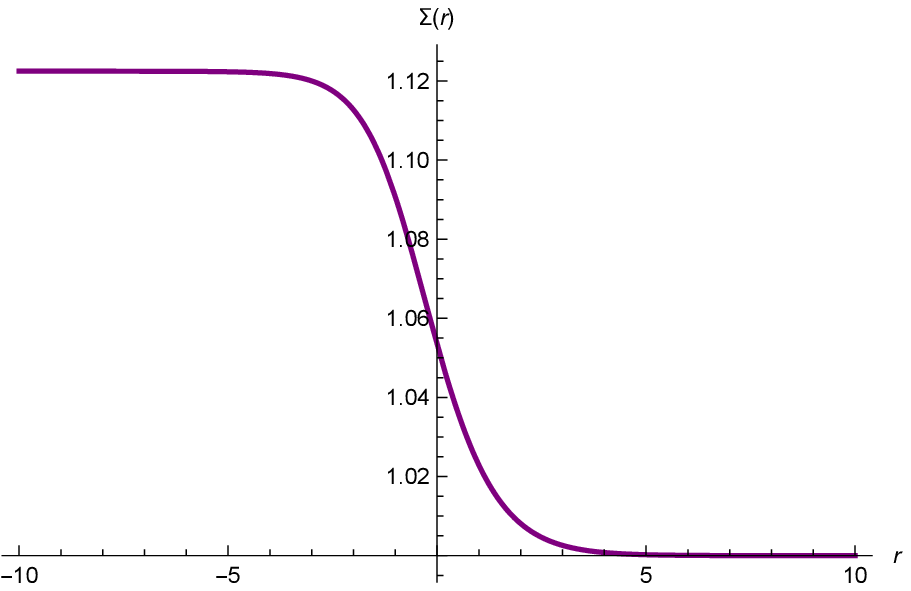}
                 \caption{Solution for $\Sigma$}
         \end{subfigure}\,\, 
\begin{subfigure}[b]{0.3\textwidth}
                 \includegraphics[width=\textwidth]{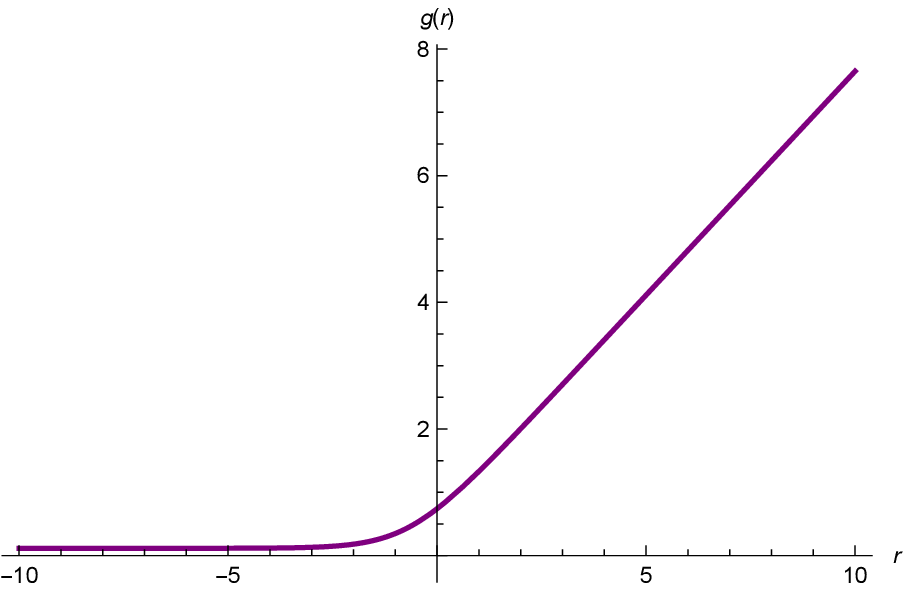}
                 \caption{Solution for $g$}
         \end{subfigure}\,\,
         ~ 
         \begin{subfigure}[b]{0.3\textwidth}
                 \includegraphics[width=\textwidth]{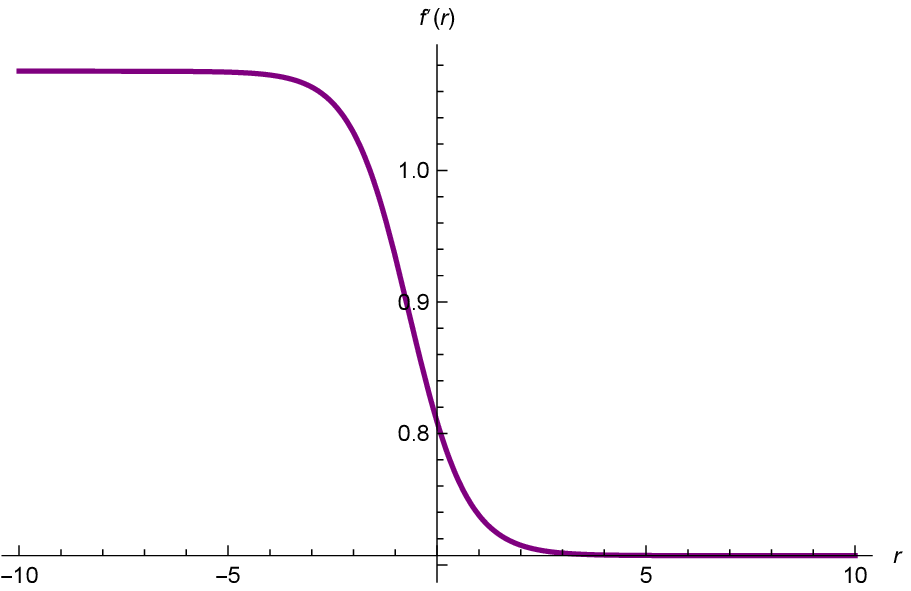}
                 \caption{Solution for $f'$}
         \end{subfigure}
         \caption{An RG flow from $N=4$ $AdS_5$ critical point with $U(1)\times SU(2)\times SU(2)$ symmetry to $N=2$ $AdS_3\times S^2$ geometry in the IR with $U(1)\times U(1)\times SU(2)$ symmetry and $g_1=1$, $a_5=1$ and $a_8=0$.}\label{fig1}
 \end{figure}

\begin{figure}
         \centering
         \begin{subfigure}[b]{0.45\textwidth}
                 \includegraphics[width=\textwidth]{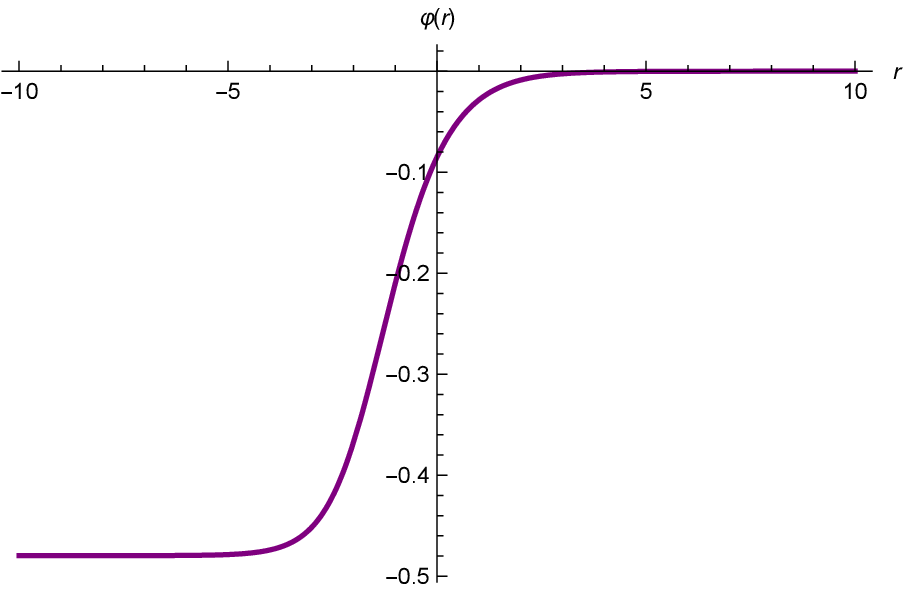}
                 \caption{Solution for $\varphi$}
         \end{subfigure} \qquad
\begin{subfigure}[b]{0.45\textwidth}
                 \includegraphics[width=\textwidth]{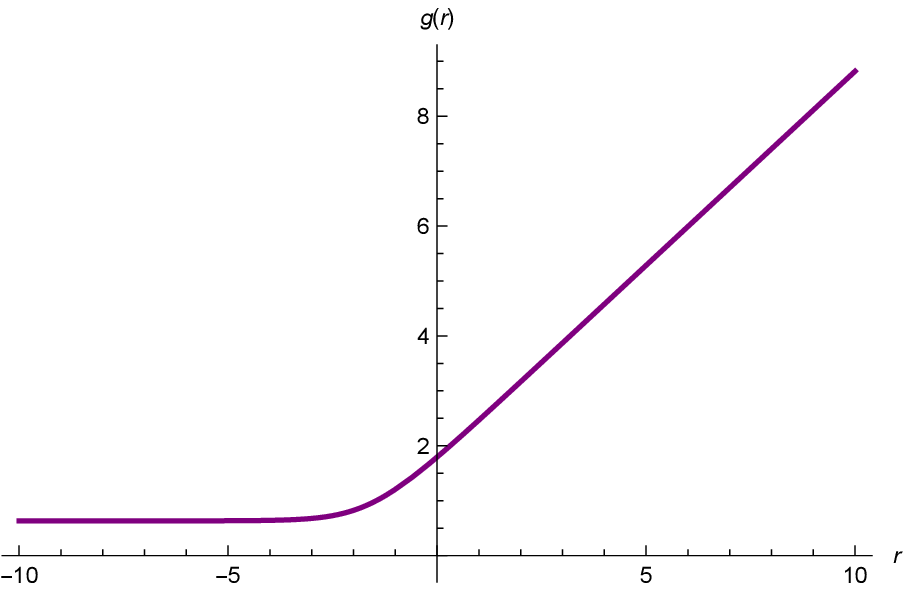}
                 \caption{Solution for $g$}
         \end{subfigure}

         ~ 
         \begin{subfigure}[b]{0.45\textwidth}
                 \includegraphics[width=\textwidth]{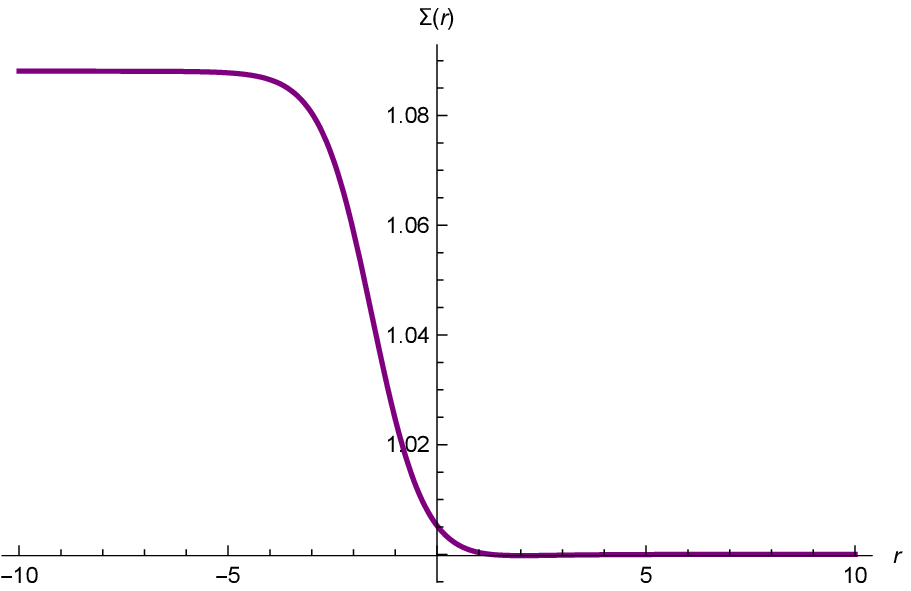}
                 \caption{Solution for $\Sigma$}
         \end{subfigure}\qquad 
         \begin{subfigure}[b]{0.45\textwidth}
                 \includegraphics[width=\textwidth]{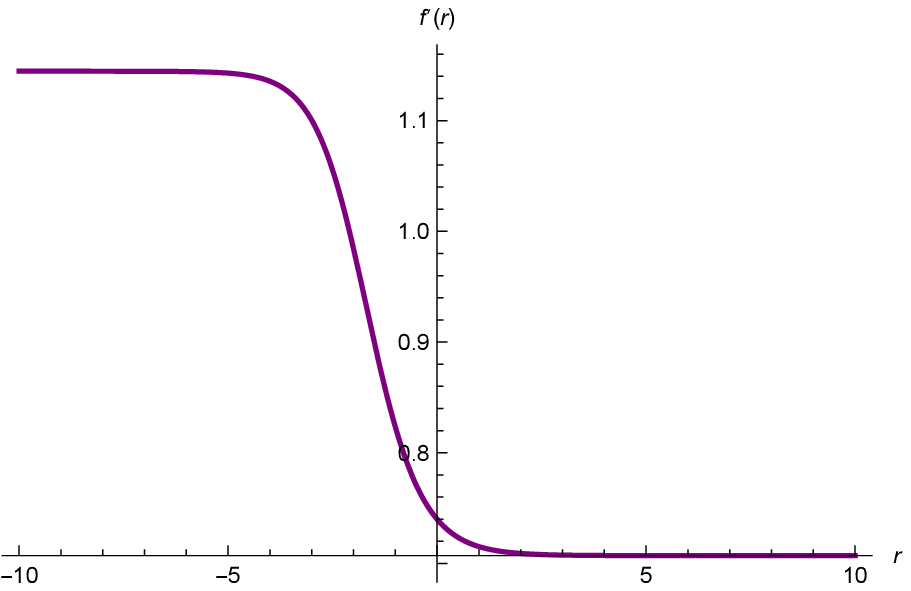}
                 \caption{Solution for $f'$}
         \end{subfigure}
         \caption{An RG flow from $N=4$ $AdS_5$ critical point with $U(1)\times SU(2)\times SU(2)$ symmetry to $N=2$ $AdS_3\times S^2$ geometry in the IR with $U(1)\times U(1)\times U(1)$ symmetry and $g_1=1$, $a_5=2$ and $a_8=-1$.}\label{fig2}
 \end{figure}

There are three singlets from the $SO(5,5)/SO(5)\times SO(5)$ coset corresponding to the $SO(5,5)$ non-compact generators $Y_{53}$, $Y_{54}$ and $Y_{55}$. However, these can be consistently truncated to only a single scalar with the coset representative given by 
\begin{equation}
\mc{V}=e^{\varphi Y_{53}}\, .\label{U1_3_coset}
\end{equation}   
\indent We now begin with the analysis for $\Sigma_2=S^2$. With the relevant component of the spin connection $\omega^{\hat{\phi}\hat{\theta}}=e^{-g}\cot \theta e^{\hat{\phi}}$, we find the covariant derivative of $\epsilon_i$ along the $\hat{\phi}$ direction    
\begin{equation}
D_{\hat{\phi}}\epsilon_i= \ldots +\frac{1}{2}e^{-g}\cot\theta\left[\gamma_{\hat{\phi}\hat{\theta}}\epsilon_i-ia_0g_1{(\sigma_2\otimes\sigma_3)_i}^j\epsilon_j+ia_5g_2{(\sigma_1\otimes \sigma_1)_i}^j\epsilon_j\right]\label{Covariant_Sigma2}
\end{equation}
where $\ldots$ refers to the term involving $g'$ that is not relevant to the present discussion. Note also that $a_8$ does not appear in the above equation since $A^8$ is not part of the R-symmetry under which the gravitini and supersymmetry parameters are charged.
\\
\indent For half-supersymmetric solutions considered in \cite{5D_N4_flow}, it has been shown that the twists from $A^0$ and $A^5$ can not be performed simultaneously, and there exist only $AdS_3\times H^2$ solutions. However, if we allow for an extra projector such that only $\frac{1}{4}$ of the original supersymmetry is unbroken, it is possible to keep both the twists from $A^0$ and $A^5$ non-vanishing. To achieve this, we note that 
\begin{equation}
i\sigma_2\otimes \sigma_3=i(\sigma_1\otimes \sigma_1)(\sigma_3\otimes \sigma_2).
\end{equation}
We then impose the following projector to make the two terms with $a_0$ and $a_5$ in \eqref{Covariant_Sigma2} proportional
\begin{equation}
{(\sigma_3\otimes \sigma_2)_i}^j\epsilon_j=-\epsilon_i\, .\label{extra_projector1}
\end{equation}
To cancel the spin connection, we then impose another projector
\begin{eqnarray}
i\gamma_{\hat{\theta}\hat{\phi}}\epsilon_i=-{(\sigma_1\otimes \sigma_1)_i}^j\epsilon_j\, .\label{theta_phi_projector1}
\end{eqnarray}
and the twist condition
\begin{equation}
a_0g_1+a_5g_2=1\, .\label{twist_con1}
\end{equation}
It should be noted that the condition \eqref{twist_con1} reduces to that of \cite{5D_N4_flow} for either $a_0=0$ or $a_5=0$. However, the solutions in this case preserve only four supercharges, or $N=2$ supersymmetry in three dimensions, due to the additional projector \eqref{extra_projector1}.
\\
\indent To setup the BPS equations, we also need the $\gamma_r$ projection due to the radial dependence of scalars. Following \cite{5D_N4_flow}, this projector is given by 
\begin{equation}
\gamma_r\epsilon_i={I_i}^j\epsilon_j\label{gamma_r_projector}
\end{equation} 
with ${I_i}^j$ defined by
\begin{equation} 
{I_i}^j={(\sigma_2\otimes \sigma_3)_i}^j\, .\label{I_gamma_r}
\end{equation}
The covariant field strength tensors for the gauge fields in \eqref{vector_ansatz_S2} can be straightforwardly computed, and the result is
\begin{equation}
\mc{H}^{\mc{M}}=-a_{\mc{M}}\sin\theta d\theta\wedge d\phi\, .
\end{equation}
\indent For $\Sigma_2=H^2$, the cancellation of the spin connection $\omega^{\hat{\phi}\hat{\theta}}=e^{-g}\coth \theta e^{\hat{\phi}}$ is again achieved by the gauge field ansatz \eqref{vector_ansatz_H2} using the conditions \eqref{extra_projector1}, \eqref{theta_phi_projector1} and \eqref{twist_con1}. On the other hand, the covariant field strengths are now given by
\begin{equation}
\mc{H}^{\mc{M}}=a_{\mc{M}}\sinh\theta d\theta\wedge d\phi\, .
\end{equation}
which have opposite signs to those of the $S^2$ case. This results in a sign change of the parameter $(a_0,a_5,a_8)$ in the corresponding BPS equations. 
\\
\indent With all these, we obtain the following BPS equations
\begin{eqnarray}
\varphi'&=&\frac{1}{2}\Sigma^{-1}e^{-\varphi-2g}\left[g_2e^{2g}(e^{2\varphi}-1)-\kappa \Sigma^2(a_5-a_8-e^{2\varphi}(a_5+a_8))\right],\\
\Sigma'&=&-\frac{1}{3}(\sqrt{2}g_1\Sigma^3+g_2\cosh\varphi)+\frac{1}{3}\Sigma^{-1}e^{-2g}[-\sqrt{2}\kappa a_0\nonumber \\
& &+\kappa\Sigma^3(a_5\cosh\varphi+a_8\sinh\varphi)],\\
g'&=&\frac{1}{6}\Sigma^{-2}\left[\sqrt{2}g_1\Sigma^4-2\sqrt{2}\kappa a_0e^{-2g}-2g_2\cosh\varphi \Sigma\right.\nonumber \\
& & \left.-4\kappa\Sigma^3e^{-2g}(a_5\cosh\varphi+a_8\sinh\varphi) \right],\\
f'&=&\frac{1}{6}\Sigma^{-2}\left[\sqrt{2}g_1\Sigma^4+\sqrt{2}\kappa a_0e^{-2g}-2g_2\cosh\varphi \Sigma \right.\nonumber \\
& & \left.+2\kappa\Sigma^3e^{-2g}(a_5\cosh\varphi+a_8\sinh\varphi) \right].
\end{eqnarray}
In these equations, $\kappa=1$ and $\kappa=-1$ refer to $\Sigma_2=S^2$ and $\Sigma_2=H^2$, respectively. It can also be readily verified that these equations also imply the second order field equations.
\\
\indent We now look for $AdS_3$ solutions from the above BPS equations. These solutions are characterized by the conditions $g'=\varphi'=\Sigma'=0$ and $f'=\frac{1}{L_{AdS_3}}$. We find the following $AdS_3$ solutions.
\begin{itemize}
\item For $\varphi=0$, $AdS_3$ solutions only exist for $a_8=0$ and are given by
\begin{equation}
\Sigma=\frac{2^{\frac{1}{6}}\kappa}{(a_5g_1)^{\frac{1}{3}}},\qquad g=\frac{1}{6}\ln\left(\frac{2a_5^4}{g_1^2}\right),\qquad L_{AdS_3}=\frac{2^{\frac{7}{6}}a_5^{\frac{2}{3}}}{g_1^{\frac{1}{3}}(1-\kappa a_5g_2)}\, .\label{AdS3_compact1}
\end{equation} 
This should be identified with similar solutions of pure $N=4$ gauged supergravity found in \cite{5D_N4_Romans}. Since $a_8$ and $\varphi$ vanish in this case, the $AdS_3$ solution has a larger symmetry $U(1)\times U(1)\times SU(2)$. Note also that unlike half-supersymmetric solutions that exist only for $\Sigma_2=H^2$, both $\Sigma_2=S^2$ and $\Sigma_2=H^2$ are possible by appropriately chosen values of $a_0$, $a_5$ and $g_1$, recall that $g_2=-\sqrt{2}g_1$. 
\item For $\varphi\neq 0$, we find a class of solutions
\begin{eqnarray}
\varphi&=&\frac{1}{2}\ln\left[\frac{(a_5-a_8)(a_0g_1-a_8g_2)}{(a_5+a_8)(a_0g_1+a_8g_2)}\right],\nonumber  \\ 
\Sigma&=&\left(\frac{\sqrt{2}\kappa a_0}{\sqrt{(a_5^2-a_8^2)(a_0^2g_1^2-a_8^2g_2^2)}}\right)^{\frac{1}{3}},\nonumber  \\
g&=&\frac{1}{6}\ln\left[\frac{2a_0^2(a_5^2-a_8^2)}{a_0^2g_1^2-a_8^2g_2^2}\right],\nonumber \\
L_{AdS_3}&=&\frac{2^{\frac{7}{6}}a_0^{\frac{1}{3}}(a_5^2-a_8^2)^{\frac{1}{3}}(a_0^2g_1^2-a_8^2g_2^2)^{\frac{1}{3}}}{a_0g_1(1-\kappa a_5g_2)-\kappa g_2^2a_8^2}\, .
\end{eqnarray}
Note that when $a_8=0$, we recover the $AdS_3$ solutions in \eqref{AdS3_compact1}. As in the previous solution, it can also be verified that these $AdS_3$ solutions exist for both $\Sigma_2=S^2$ and $\Sigma_2=H^2$.
\end{itemize}
\indent Examples of numerical solutions interpolating between $N=4$ $AdS_5$ vacuum with $U(1)\times SU(2)\times SU(2)$ symmetry to these $AdS_3\times \Sigma_2$ are shown in figure \ref{fig1} and \ref{fig2}. At large $r$, the solutions are asymptotically $N=4$ supersymmetric $AdS_5$ critical point i given in \eqref{AdS5_1_compact}. It should also be noted that the flow solutions preserve only two supercharges due to the $\gamma_r$ projector imposed along the flow. 

\subsubsection{Solutions with $U(1)\times U(1)_{\textrm{diag}}$ symmetry}  
\begin{figure}
         \centering
         \begin{subfigure}[b]{0.45\textwidth}
                 \includegraphics[width=\textwidth]{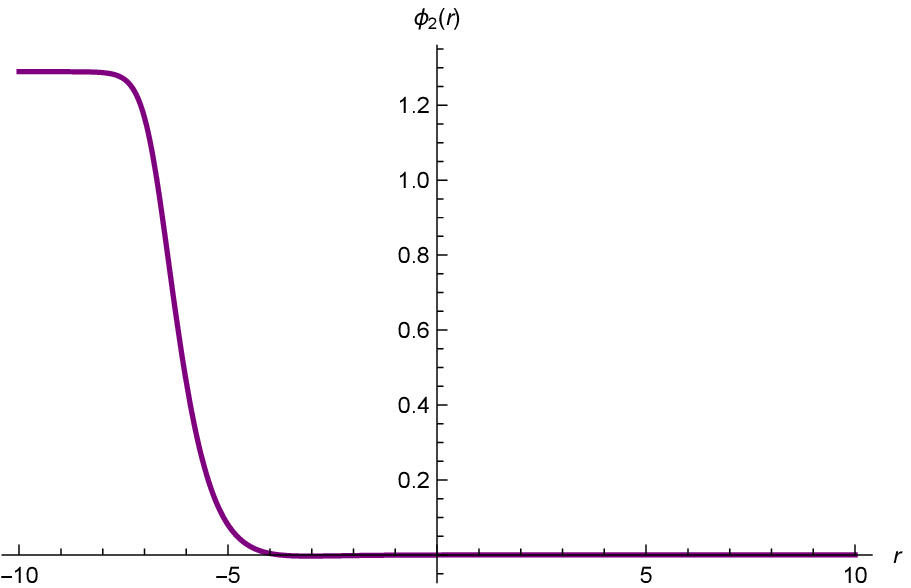}
                 \caption{Solution for $\phi_2$}
         \end{subfigure} \qquad 
\begin{subfigure}[b]{0.45\textwidth}
                 \includegraphics[width=\textwidth]{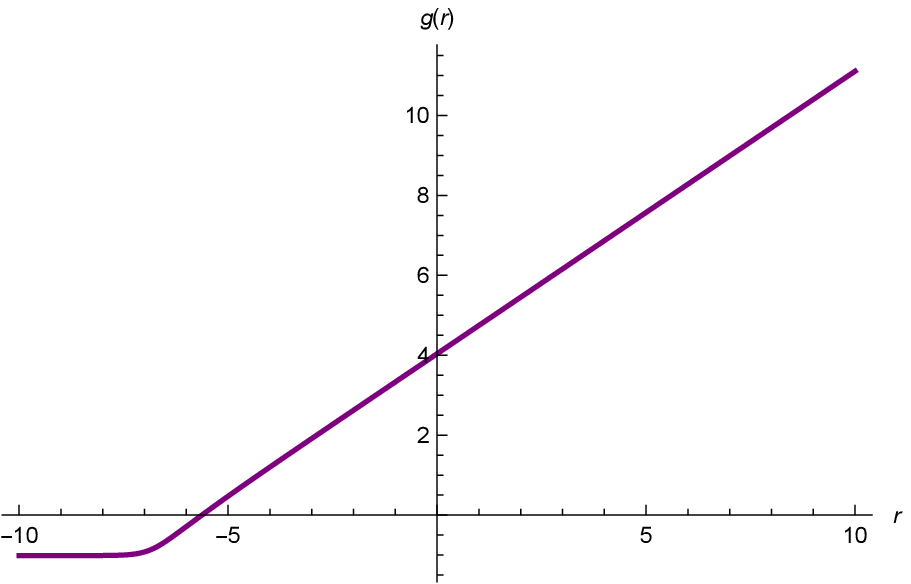}
                 \caption{Solution for $g$}
         \end{subfigure}\\

         ~ 
         \begin{subfigure}[b]{0.45\textwidth}
                 \includegraphics[width=\textwidth]{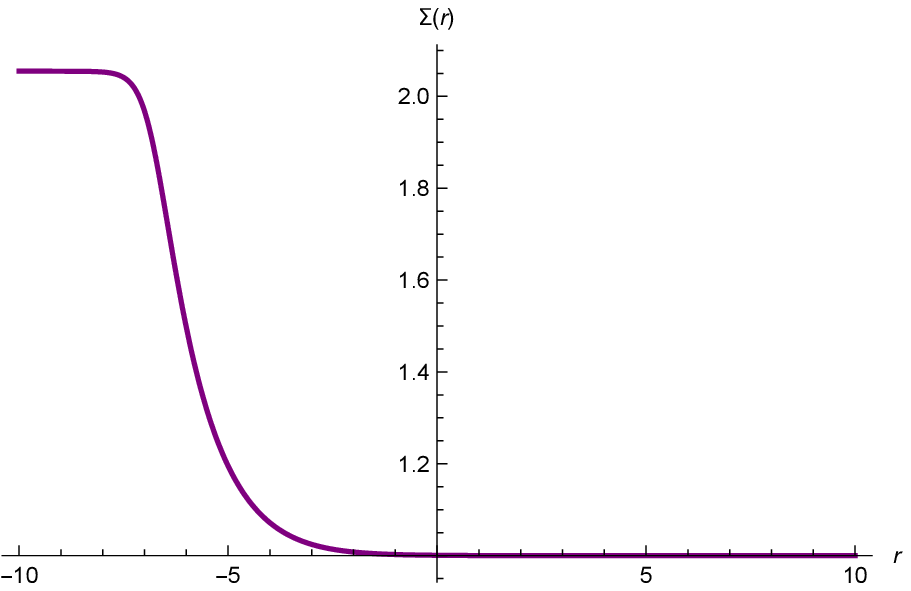}
                 \caption{Solution for $\Sigma$}
         \end{subfigure}\qquad 
         \begin{subfigure}[b]{0.45\textwidth}
                 \includegraphics[width=\textwidth]{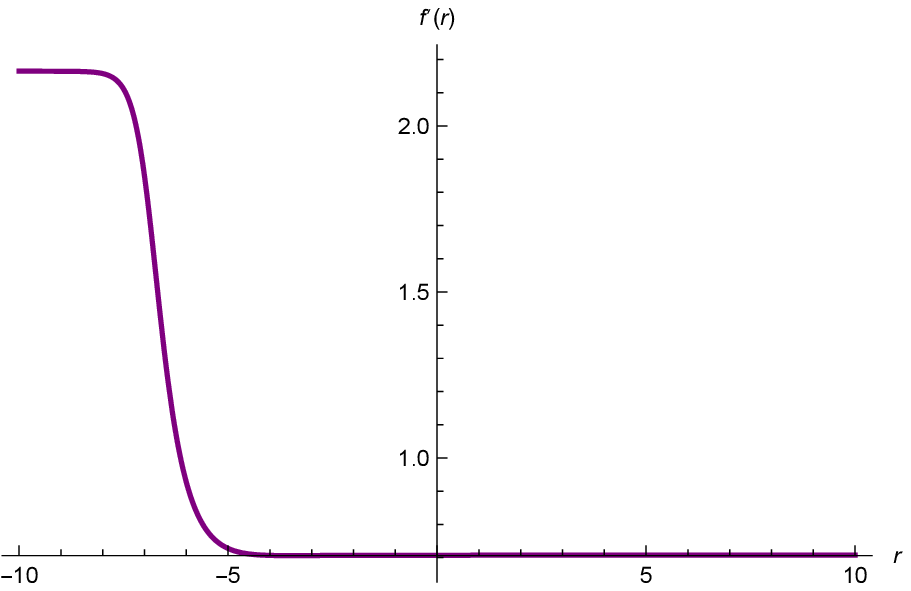}
                 \caption{Solution for $f'$}
         \end{subfigure}
         \caption{An RG flow from $AdS_5$ critical point with $U(1)\times SU(2)\times SU(2)$ symmetry to $AdS_3\times S^2$ critical point I for $g_1=1$, $g_3=2g_1$ and $a_5=\frac{1}{4}$.}\label{fig3}
\end{figure}

\begin{figure}
         \centering
         \begin{subfigure}[b]{0.3\textwidth}
                 \includegraphics[width=\textwidth]{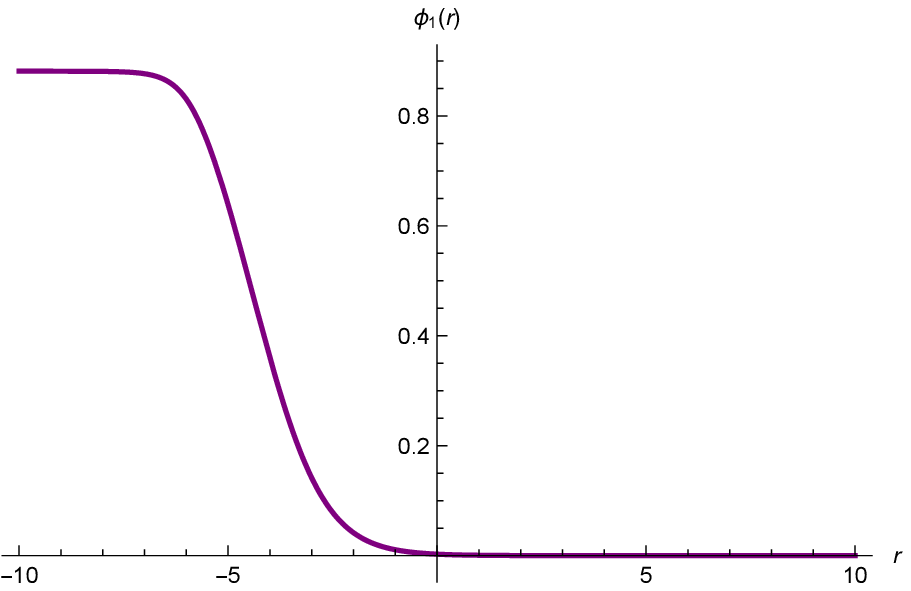}
                 \caption{Solution for $\phi_1$}
         \end{subfigure} \,\,\,
\begin{subfigure}[b]{0.3\textwidth}
                 \includegraphics[width=\textwidth]{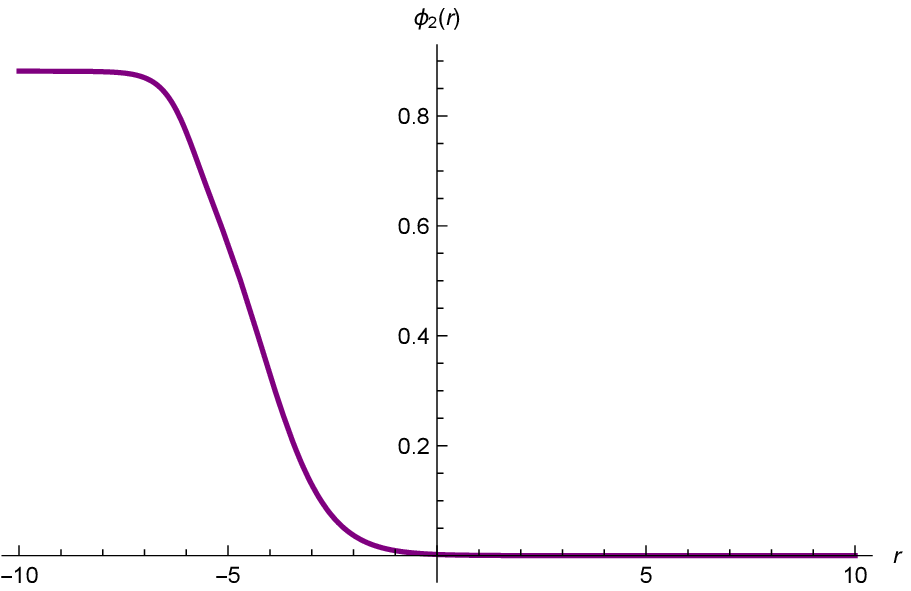}
                 \caption{Solution for $\phi_2$}
         \end{subfigure}\,\,\,
\begin{subfigure}[b]{0.3\textwidth}
                 \includegraphics[width=\textwidth]{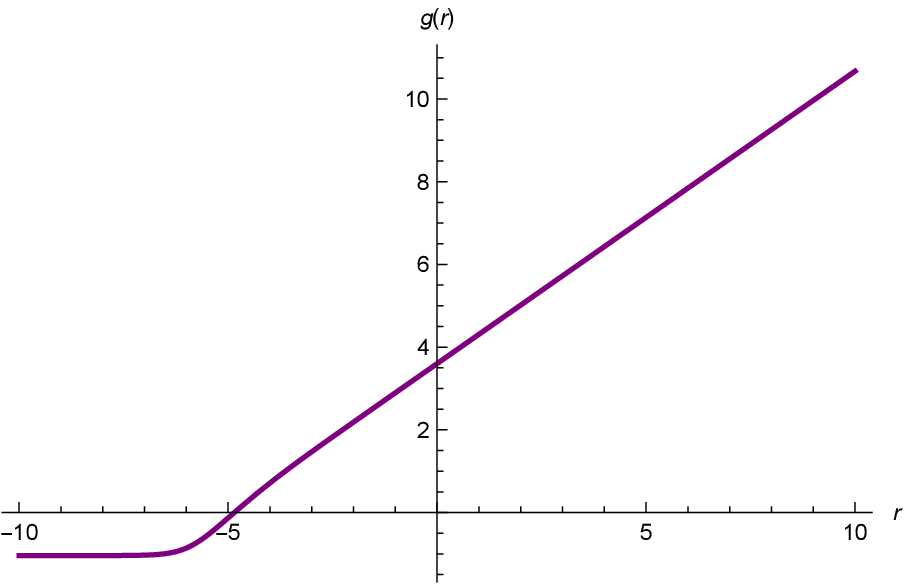}
                 \caption{Solution for $g$}
         \end{subfigure}\\

         ~ 
         \begin{subfigure}[b]{0.45\textwidth}
                 \includegraphics[width=\textwidth]{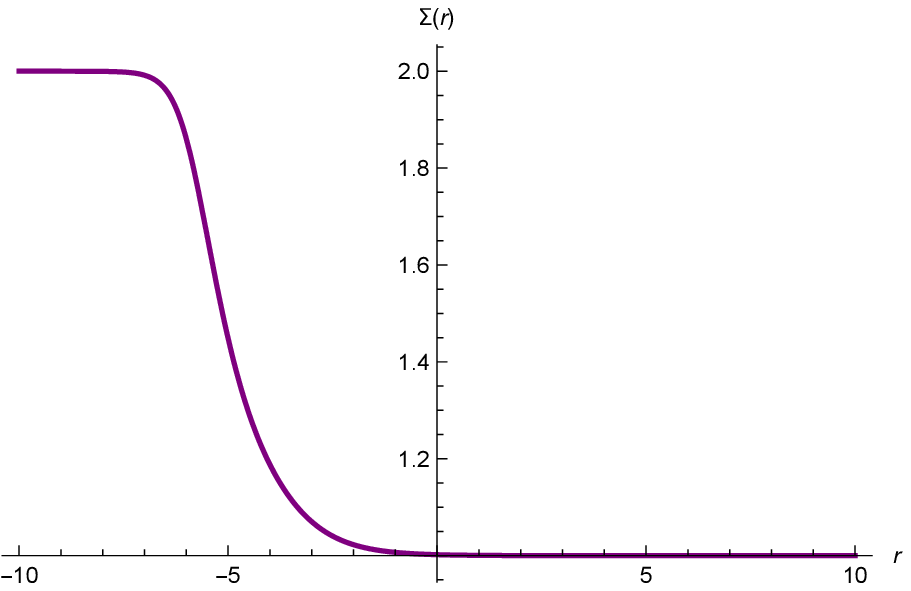}
                 \caption{Solution for $\Sigma$}
         \end{subfigure}\qquad 
         \begin{subfigure}[b]{0.45\textwidth}
                 \includegraphics[width=\textwidth]{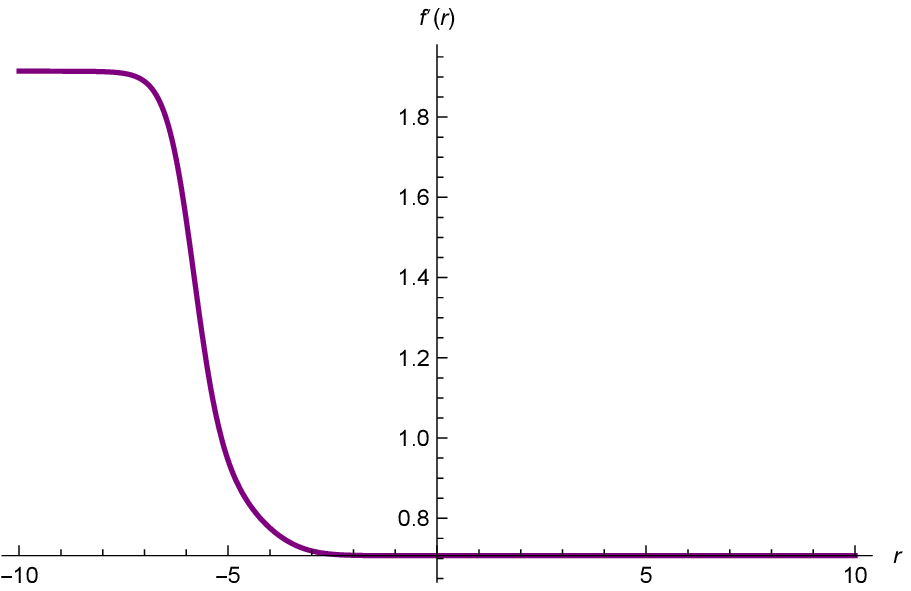}
                 \caption{Solution for $f'$}
         \end{subfigure}
         \caption{An RG flow from $AdS_5$ critical point with $U(1)\times SU(2)\times SU(2)$ symmetry to $AdS_3\times S^2$ critical point II for $g_1=1$, $g_3=2g_1$ and $a_5=\frac{1}{4}$.}\label{fig4}
 \end{figure}

\begin{figure}
         \centering
         \begin{subfigure}[b]{0.3\textwidth}
                 \includegraphics[width=\textwidth]{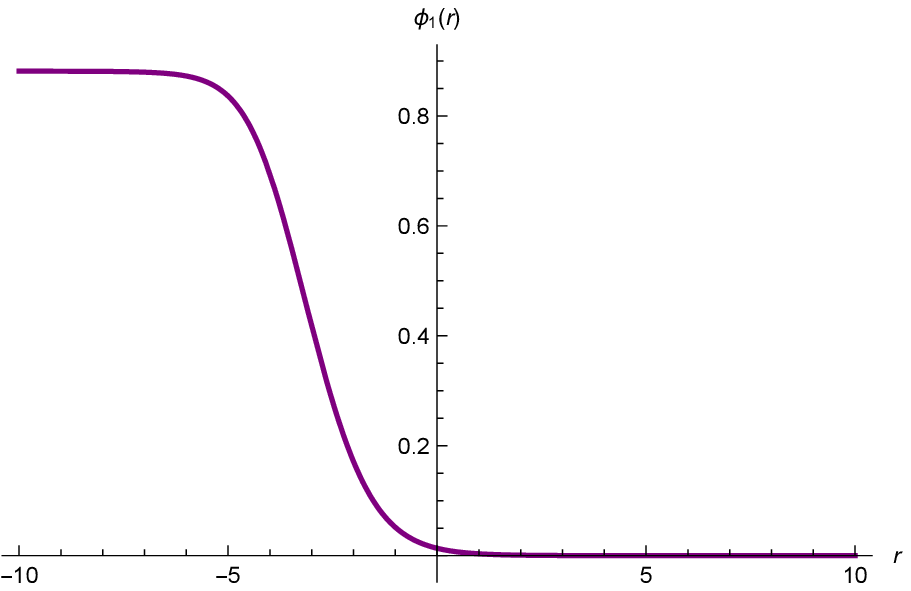}
                 \caption{Solution for $\phi_1$}
         \end{subfigure} \,\,\,
\begin{subfigure}[b]{0.3\textwidth}
                 \includegraphics[width=\textwidth]{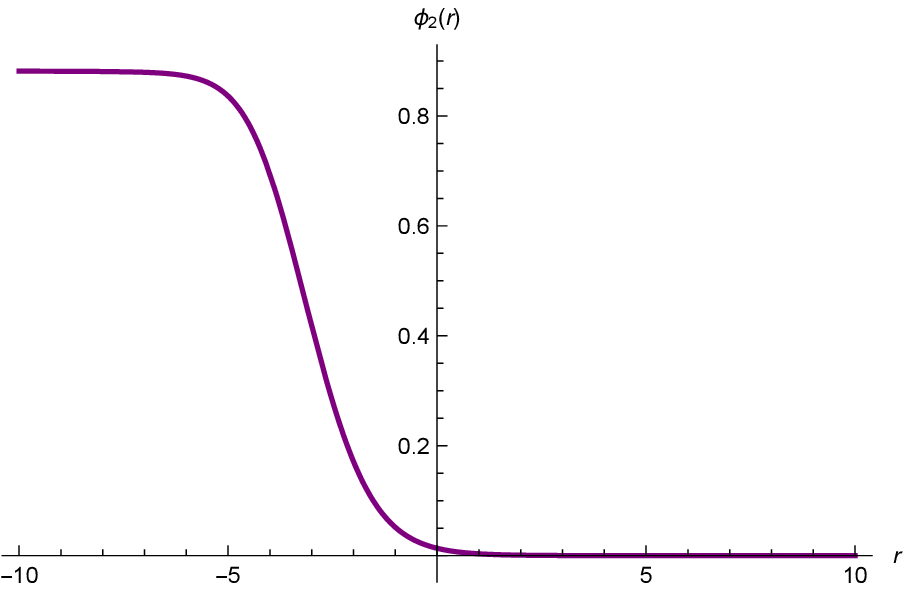}
                 \caption{Solution for $\phi_2$}
         \end{subfigure}\,\,\,
\begin{subfigure}[b]{0.3\textwidth}
                 \includegraphics[width=\textwidth]{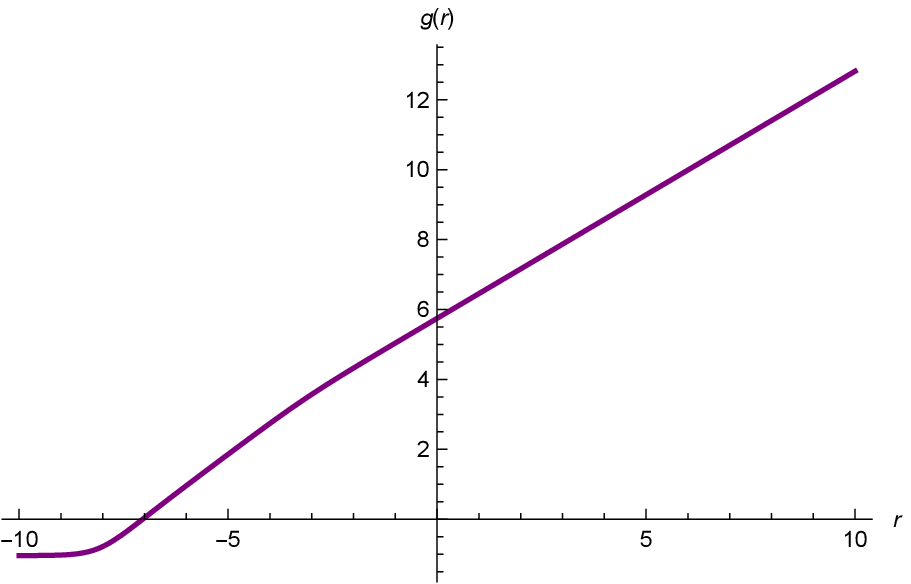}
                 \caption{Solution for $g$}
         \end{subfigure}\\

         ~ 
         \begin{subfigure}[b]{0.45\textwidth}
                 \includegraphics[width=\textwidth]{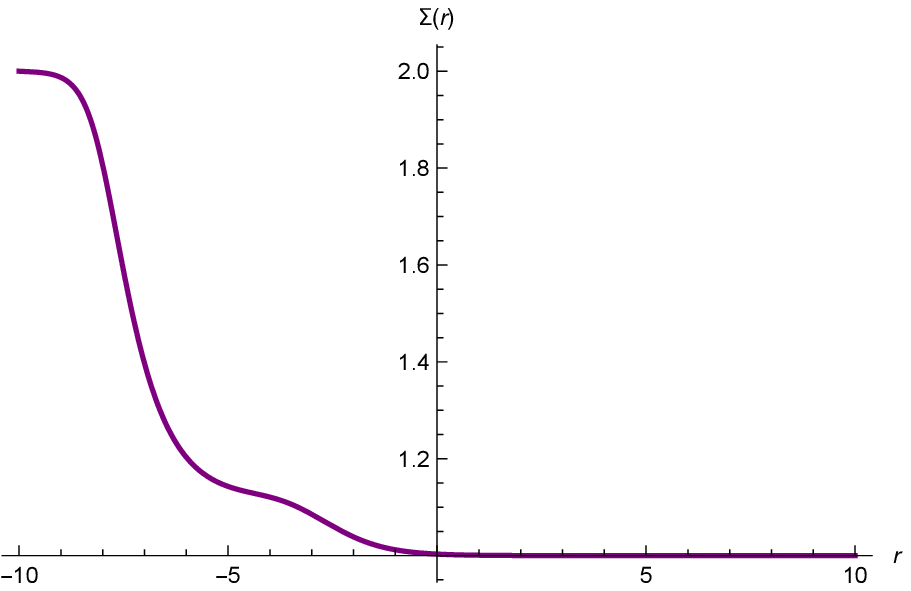}
                 \caption{Solution for $\Sigma$}
         \end{subfigure}\qquad 
         \begin{subfigure}[b]{0.45\textwidth}
                 \includegraphics[width=\textwidth]{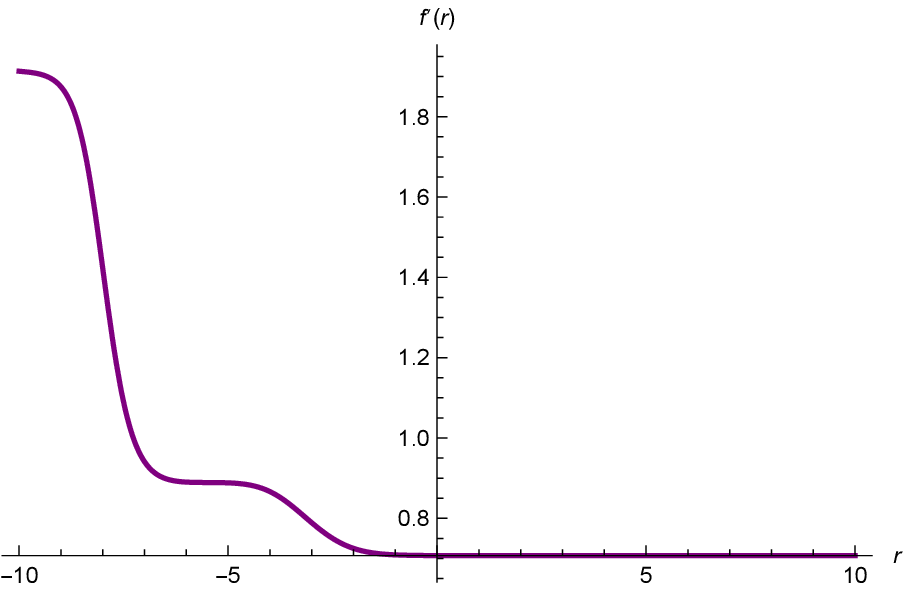}
                 \caption{Solution for $f'$}
         \end{subfigure}
         \caption{An RG flow from $AdS_5$ critical point with $U(1)\times SU(2)\times SU(2)$ symmetry to $AdS_5$ critical point with $U(1)\times SU(2)_{\textrm{diag}}$ symmetry and finally to $AdS_3\times S^2$ critical point II for $g_1=1$, $g_3=2g_1$ and $a_5=\frac{1}{4}$.}\label{fig5}
 \end{figure}

\begin{figure}
         \centering
         \begin{subfigure}[b]{0.3\textwidth}
                 \includegraphics[width=\textwidth]{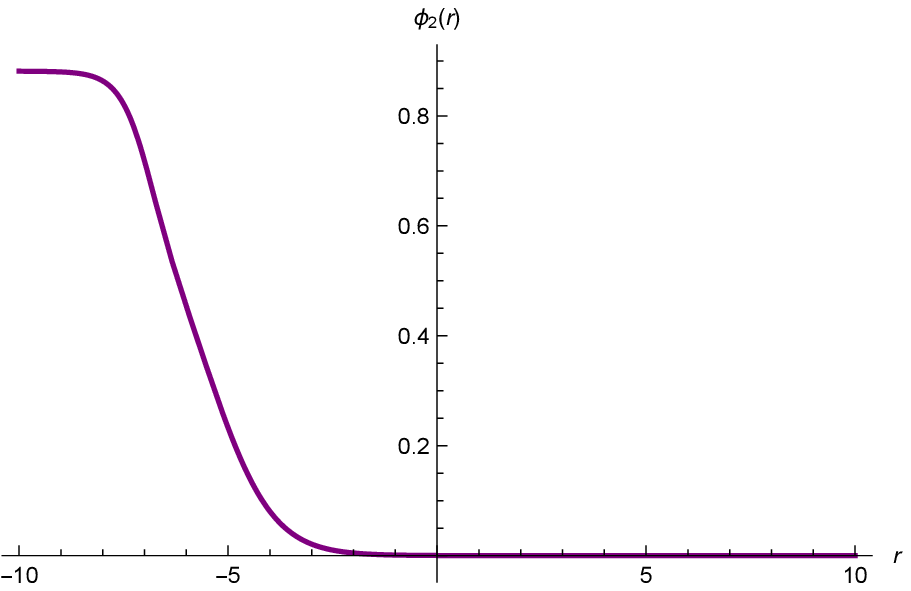}
                 \caption{Solution for $\phi_2$}
         \end{subfigure} \,\,\,
\begin{subfigure}[b]{0.3\textwidth}
                 \includegraphics[width=\textwidth]{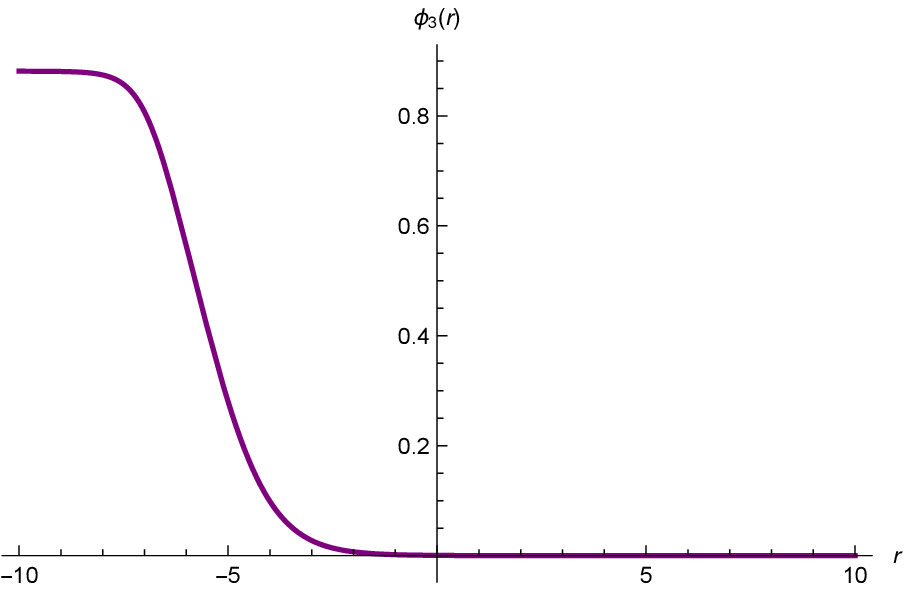}
                 \caption{Solution for $\phi_3$}
         \end{subfigure}\,\,\,
\begin{subfigure}[b]{0.3\textwidth}
                 \includegraphics[width=\textwidth]{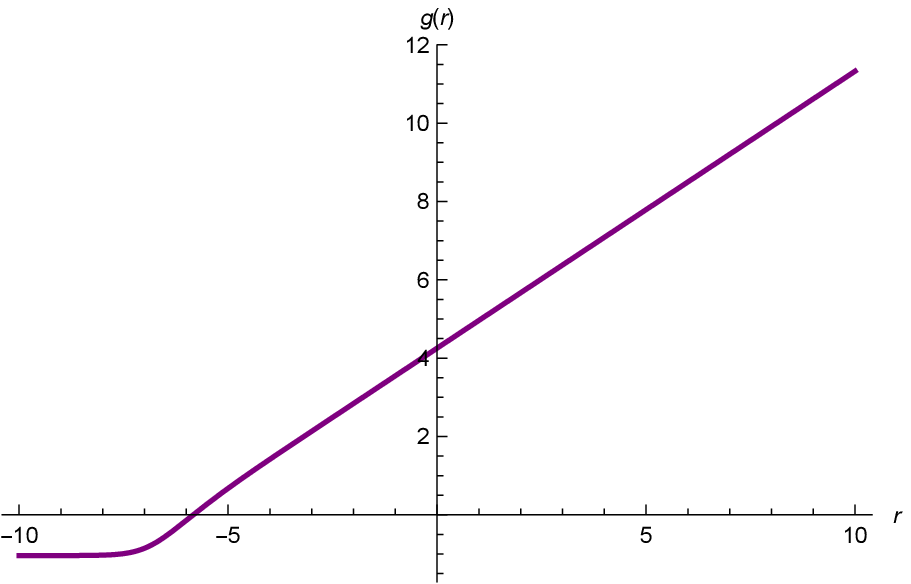}
                 \caption{Solution for $g$}
         \end{subfigure}\\

         ~ 
         \begin{subfigure}[b]{0.45\textwidth}
                 \includegraphics[width=\textwidth]{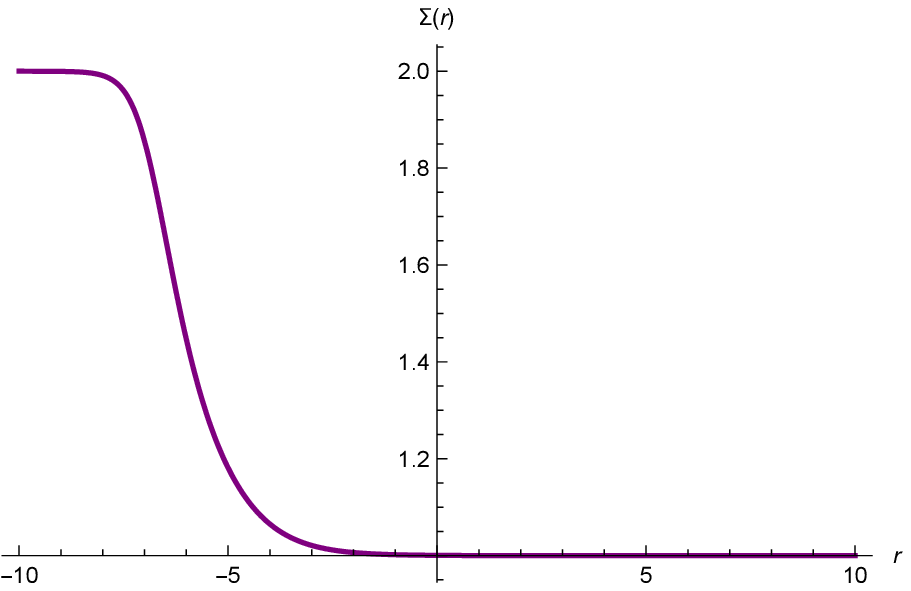}
                 \caption{Solution for $\Sigma$}
         \end{subfigure}\qquad 
         \begin{subfigure}[b]{0.45\textwidth}
                 \includegraphics[width=\textwidth]{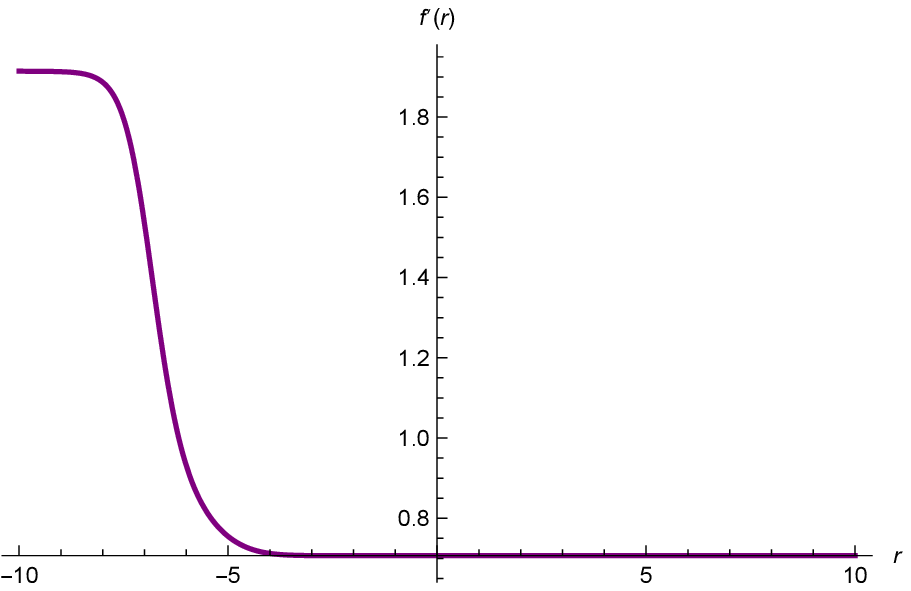}
                 \caption{Solution for $f'$}
         \end{subfigure}
         \caption{An RG flow from $AdS_5$ critical point with $U(1)\times SU(2)\times SU(2)$ symmetry to $AdS_3\times S^2$ critical point III for $g_1=1$, $g_3=2g_1$ and $a_5=\frac{1}{4}$.}\label{fig6}
\end{figure}

We now move to a set of scalars with smaller unbroken symmetry $U(1)\times U(1)_{\textrm{diag}}$ with $U(1)_{\textrm{diag}}$ being a diagonal subgroup of $U(1)\times U(1)\subset SU(2)\times SU(2)$. As pointed out in \cite{5D_N4_flow}, there are five singlets from the vector multiplet scalars but these can be truncated to three scalars corresponding to the following non-compact generators of $SO(5,5)$
\begin{equation}
\hat{Y}_1=Y_{31}+Y_{42},\qquad \hat{Y}_2=Y_{53},\qquad \hat{Y}_3=Y_{32}-Y_{41}\, .
\end{equation}
The coset representative is then given by
\begin{equation}
\mc{V}=e^{\phi_1\hat{Y}_1}e^{\phi_2\hat{Y}_2}e^{\phi_3\hat{Y}_3}\, .\label{U1_U1d_coset}
\end{equation}
To implement the $U(1)_{\textrm{diag}}$ gauge symmetry, we impose an additional condition on the parameters $a_5$ and $a_8$ as follow
\begin{equation}
g_2a_5=g_3a_8\, .
\end{equation}
We can repeat the previous analysis for the $U(1)\times U(1)\times U(1)$ twists, and the result is the same as in the previous case with the twist condition \eqref{twist_con1} and projectors \eqref{extra_projector1}, \eqref{theta_phi_projector1} and \eqref{gamma_r_projector}.
\\
\indent With the same procedure as in the previous case, we obtain the following BPS equations
\begin{eqnarray}
\phi_1'&=&\frac{1}{2}\Sigma^{-1}\textrm{sech}(2\phi_3)\sinh(2\phi_1)(g_2\cosh\phi_2+g_3\sinh\phi_2),\\
\phi_2'&=&\frac{1}{2}\Sigma^{-1}\cosh(2\phi_1)\cosh(2\phi_3)(g_2\sinh\phi_2+g_3\cosh\phi_2)\nonumber \\
& &+\frac{1}{2}\Sigma^{-1}(g_2\sinh\phi_2-g_3\cosh\phi_2)+\frac{a_5\kappa}{g_3}e^{-2g}\Sigma(g_2\cosh\phi_2+g_3\sinh\phi_2),\quad\\
\phi_3'&=&\frac{1}{2}\Sigma^{-1}\cosh(2\phi_1)\sinh(2\phi_3)(g_2\cosh\phi_2+g_3\sinh\phi_2),\\
\Sigma'&=&-\frac{1}{6g_3}\Sigma^{-1}e^{-2g}\left[-2\kappa a_5\Sigma^3(g_3\cosh\phi_2+g_2\sinh\phi_2)+2\sqrt{2}\kappa g_3a_0 \right. \nonumber \\
& &+e^{2g}g_3\Sigma \left[\cosh(2\phi_1)\cosh(2\phi_3)(g_2\cosh\phi_2+g_3\sinh\phi_2) \right. \nonumber \\
& &\left. \left. g_2\cosh\phi_2-g_3\sinh\phi_2+2\sqrt{2}g_1\Sigma^3\right]\right],\\
g'&=&\frac{1}{6g_3}\Sigma^{-2}\left[g_3\Sigma(g_3\sinh\phi_2-g_2\cosh\phi_2)-2\sqrt{2}\kappa a_0g_3e^{-2g} \right.\nonumber \\
& &-\Sigma \cosh(2\phi_1)\cosh(2\phi_3)(g_2\cosh\phi_2+g_3\sinh\phi_2)+\sqrt{2}g_1g_3\Sigma^4\nonumber \\
& & \left. -4\kappa a_5e^{-2g}\Sigma^3(g_3\cosh\phi_2+g_2\sinh\phi_2)\right],\\
f'&=&\frac{1}{6g_3}\Sigma^{-2}\left[g_3\Sigma(g_3\sinh\phi_2-g_2\cosh\phi_2)+\sqrt{2}\kappa a_0g_3e^{-2g} \right.\nonumber \\
& &-\Sigma \cosh(2\phi_1)\cosh(2\phi_3)(g_2\cosh\phi_2+g_3\sinh\phi_2)+\sqrt{2}g_1g_3\Sigma^4\nonumber \\
& & \left. +2\kappa a_5e^{-2g}\Sigma^3(g_3\cosh\phi_2+g_2\sinh\phi_2)\right].
\end{eqnarray}
\indent From these equations, we find the following $AdS_3\times \Sigma_2$ solutions.
\begin{itemize}
\item For $\phi_1=\phi_3=0$, there is a family of $AdS_3$ solutions given by
\begin{eqnarray}
\textrm{I}: \quad \phi_2&=&\frac{1}{2}\ln\left[\frac{(g_2-g_3)(g_2^2a_5-a_0g_1g_3)}{(g_2+g_3)(g_2^2a_5+a_0g_1g_3)}\right],\nonumber \\
 g&=&\frac{1}{6}\ln\left[\frac{2a_0^2a_5^4(g_3^2-g_2^2)^2}{g_3^2(a_0^2g_1^2g_3^2-a_5^2g_2^4)}\right],\nonumber \\
\Sigma&=&-\left[\frac{\sqrt{2}a_0g_3^2}{a_5\sqrt{(g_3^2-g_2^2)(a_0^2g_1^2g_3^2-a_5^2g_2^4)}}\right]^{\frac{1}{3}}.
\end{eqnarray}
We refrain from giving the explicit form of $L_{AdS_3}$ at this vacuum due to its complexity.   
\item For $\phi_3=0$, we find
\begin{eqnarray}
\textrm{II}:\quad \phi_2&=&\phi_1=\frac{1}{2}\ln\left[\frac{g_3-g_2}{g_3+g_2}\right],\qquad \Sigma=\left[\frac{\sqrt{2}\kappa g_3}{g_1a_5\sqrt{g_3^2-g_2^2}}\right]^{\frac{1}{3}},\nonumber \\
g&=&\frac{1}{6}\ln \left[\frac{2a_5^4(g_3^2-g_2^2)^2}{g_1^2g_3^4}\right],\quad L_{AdS_3}=\left[\frac{8\sqrt{2}a_5^2(g_3^2-g_2^2)}{g_1g_3^2(1-\kappa a_5g_2)^3}\right]^{\frac{1}{3}}.\quad
\end{eqnarray}
\item Finally, for $\phi_1=0$, we find
\begin{eqnarray}
\textrm{III}:\quad \phi_2&=&\phi_3=\frac{1}{2}\ln\left[\frac{g_3-g_2}{g_3+g_2}\right],\qquad \Sigma=\left[\frac{\sqrt{2}\kappa g_3}{g_1a_5\sqrt{g_3^2-g_2^2}}\right]^{\frac{1}{3}},\nonumber \\
g&=&\frac{1}{6}\ln \left[\frac{2a_5^4(g_3^2-g_2^2)^2}{g_1^2g_3^4}\right],\quad L_{AdS_3}=\left[\frac{8\sqrt{2}a_5^2(g_3^2-g_2^2)}{g_1g_3^2(1-\kappa a_5g_2)^3}\right]^{\frac{1}{3}}. \quad
\end{eqnarray}
\end{itemize}
Unlike the previous case, at large $r$, we find that solutions to these BPS equations can be asymptotic to any of the two $N=4$ supersymmetric $AdS_5$ vacua i and ii given in \eqref{AdS5_1_compact} and \eqref{AdS5_2_compact}. Therefore, we can have RG flows from the two $AdS_5$ vacua to any of these $AdS_3\times \Sigma_2$ solutions. Some examples of these solutions for $\Sigma_2=S^2$ are given in figures \ref{fig3}, \ref{fig4}, \ref{fig5} and \ref{fig6}.    

\subsection{Supersymmetric black holes}
\begin{figure}
         \centering
         \begin{subfigure}[b]{0.45\textwidth}
                 \includegraphics[width=\textwidth]{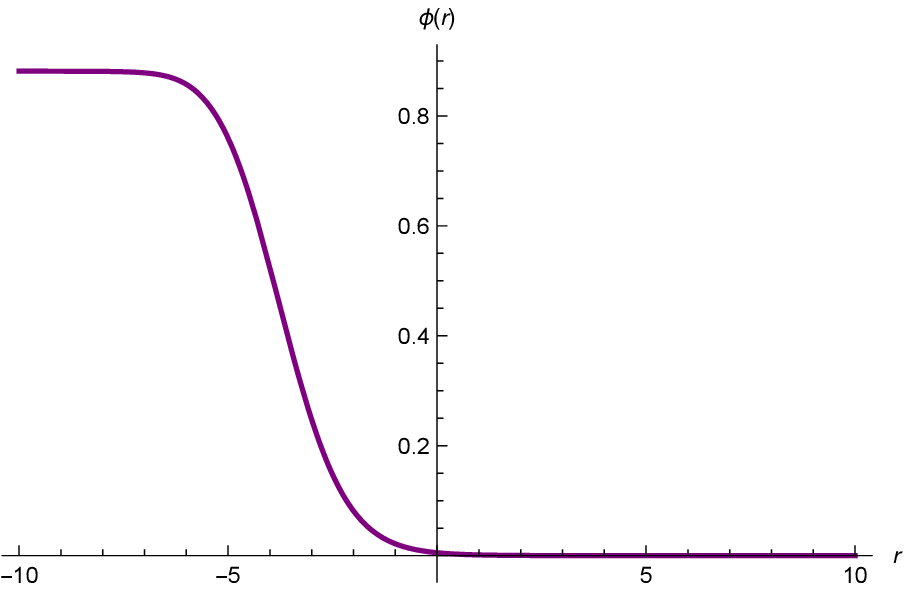}
                 \caption{Solution for $\phi$}
         \end{subfigure} \qquad 
\begin{subfigure}[b]{0.45\textwidth}
                 \includegraphics[width=\textwidth]{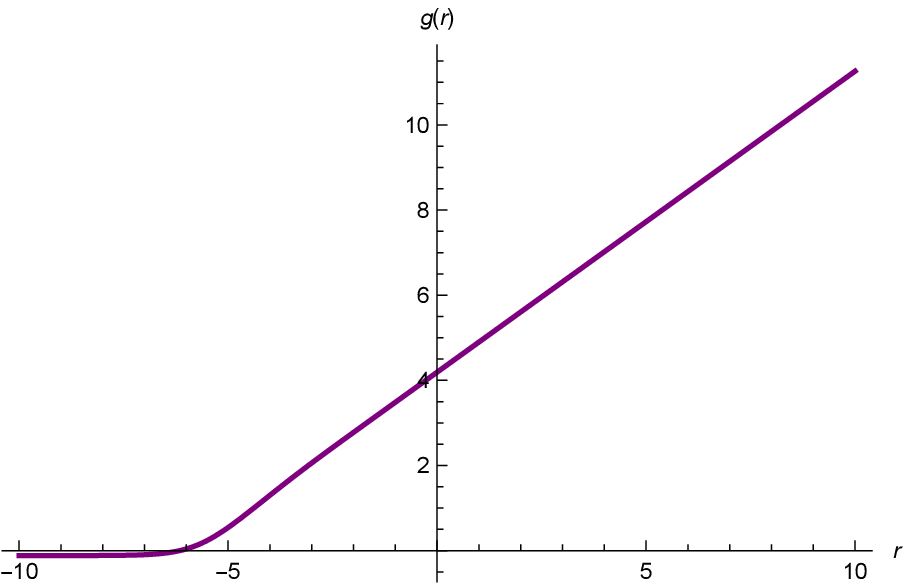}
                 \caption{Solution for $g$}
         \end{subfigure}\\

         ~ 
         \begin{subfigure}[b]{0.45\textwidth}
                 \includegraphics[width=\textwidth]{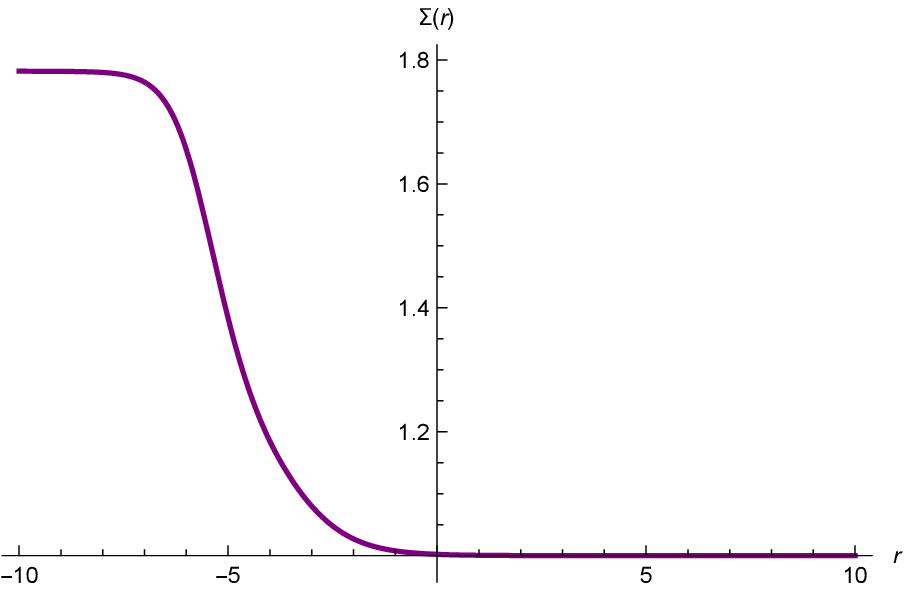}
                 \caption{Solution for $\Sigma$}
         \end{subfigure}\qquad 
         \begin{subfigure}[b]{0.45\textwidth}
                 \includegraphics[width=\textwidth]{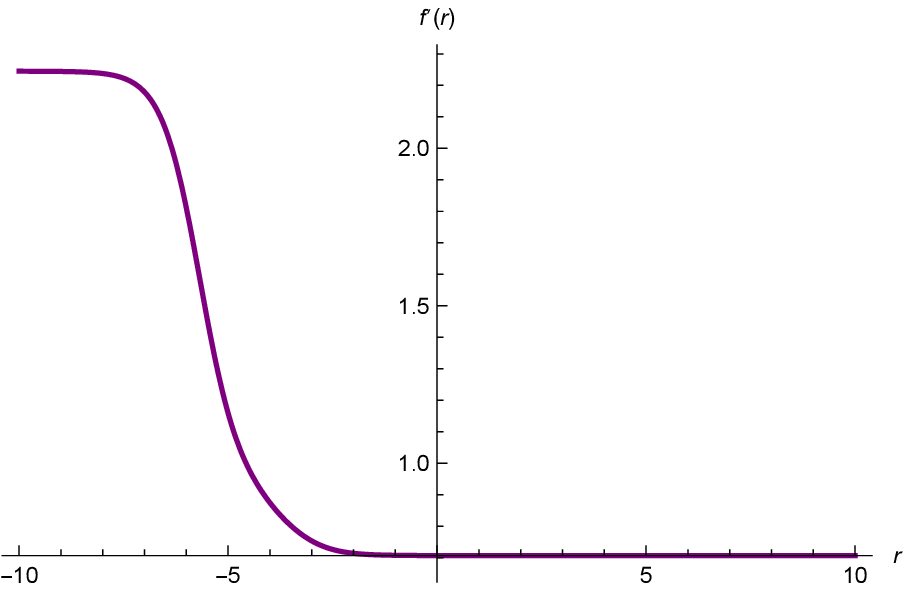}
                 \caption{Solution for $f'$}
         \end{subfigure}
         \caption{An RG flow from $AdS_5$ critical point with $U(1)\times SU(2)\times SU(2)$ symmetry to $AdS_2\times H^3$ critical point for $g_1=1$ and $g_3=2g_1$.}\label{fig7}
\end{figure}

\begin{figure}
         \centering
         \begin{subfigure}[b]{0.45\textwidth}
                 \includegraphics[width=\textwidth]{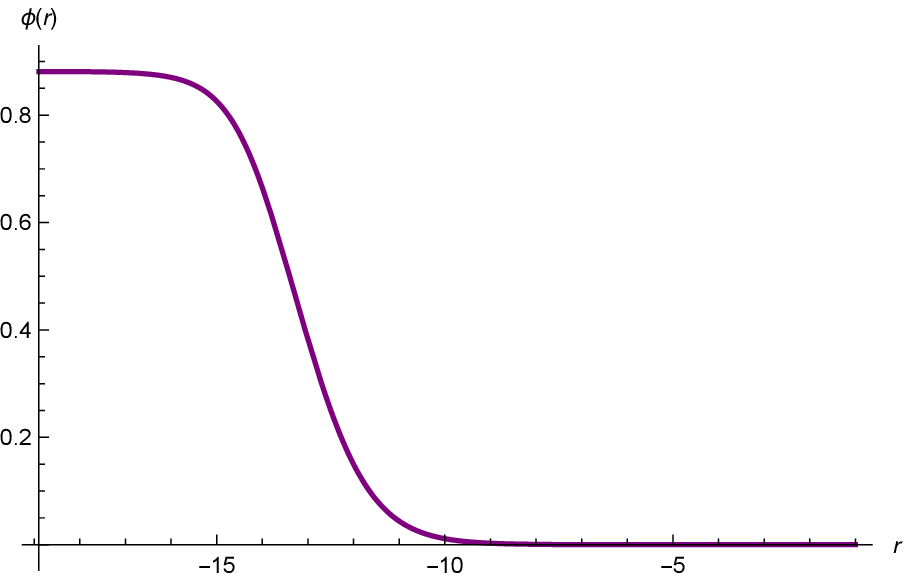}
                 \caption{Solution for $\phi$}
         \end{subfigure} \qquad
\begin{subfigure}[b]{0.45\textwidth}
                 \includegraphics[width=\textwidth]{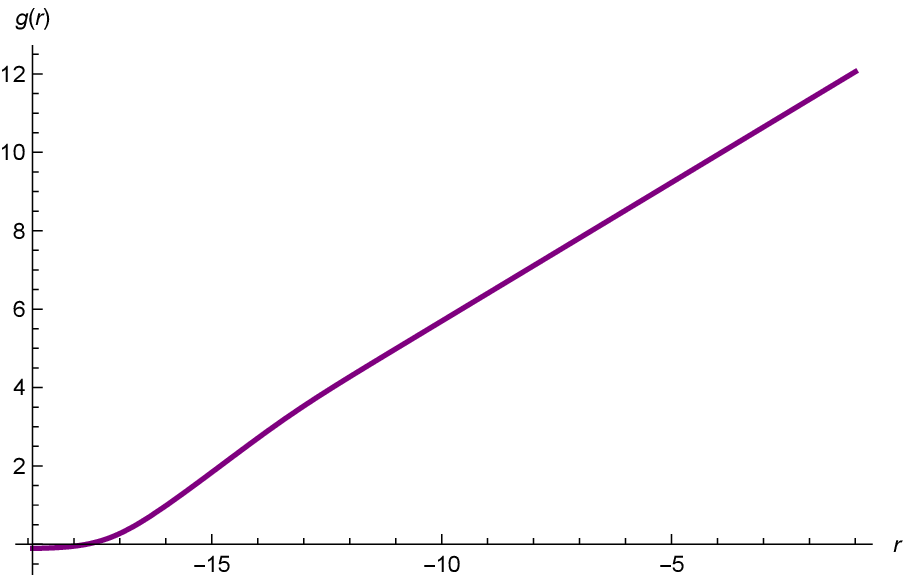}
                 \caption{Solution for $g$}
         \end{subfigure}\\

         ~ 
         \begin{subfigure}[b]{0.45\textwidth}
                 \includegraphics[width=\textwidth]{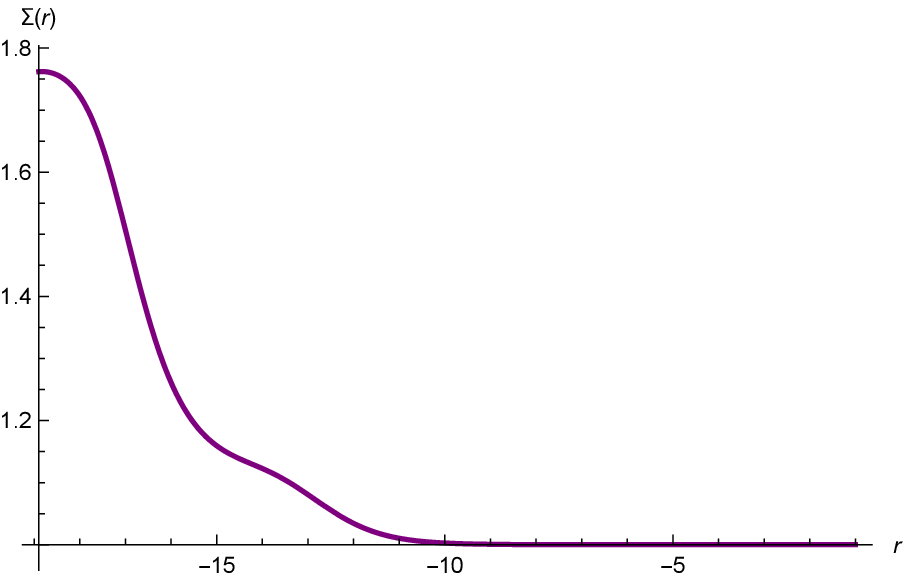}
                 \caption{Solution for $\Sigma$}
         \end{subfigure}\qquad 
         \begin{subfigure}[b]{0.45\textwidth}
                 \includegraphics[width=\textwidth]{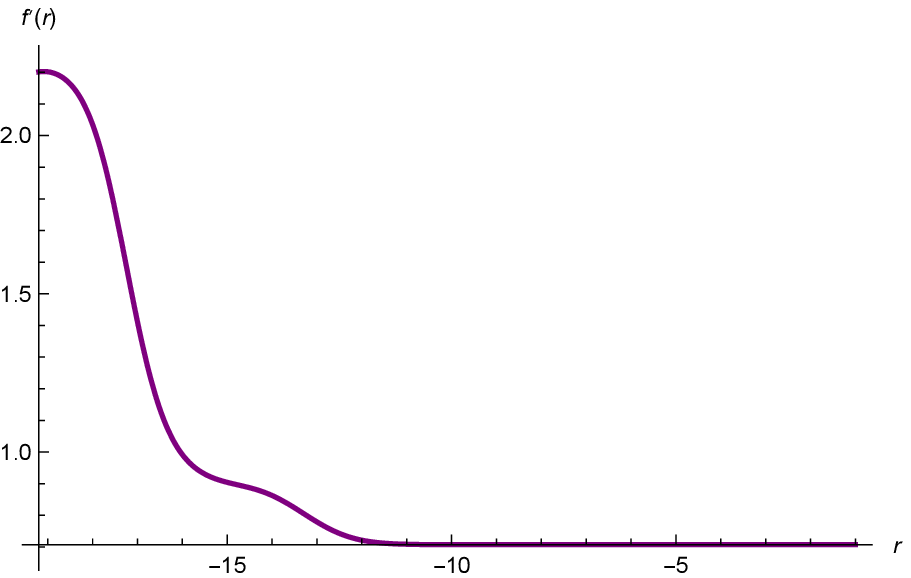}
                 \caption{Solution for $f'$}
         \end{subfigure}
         \caption{An RG flow from $AdS_5$ critical point with $U(1)\times SU(2)\times SU(2)$ symmetry to $AdS_5$ critical point with $U(1)\times SU(2)_{\textrm{diag}}$ symmetry and finally to $AdS_2\times H^3$ critical point for $g_1=1$ and $g_3=2g_1$.}\label{fig8}
 \end{figure}

We now move to another type of solutions, supersymmetric $AdS_5$ black holes. We will consider near horizon geometries of the form $AdS_2\times \Sigma_3$ for $\Sigma_3=S^3$ and $\Sigma_3=H^3$. The twist procedure is still essential to preserve supersymmetry. For the $S^3$ case, we take the metric to be
\begin{equation}
ds^2 = -e^{2f(r)}dt^2 + dr^2 + e^{2g(r)}\left[d\psi^2 + \sin^2\psi (d\theta^2 + \sin^2\theta d\phi^2)\right].\label{metric_S3}
\end{equation} 
With the following choice of vielbein
\begin{eqnarray}
e^{\hat{t}}&=&e^fdt,\qquad e^{\hat{r}}=dr,\qquad e^{\hat{\psi}}=e^gd\psi,\nonumber \\
e^{\hat{\theta}}&=&e^g\sin\psi d\theta,\qquad e^{\hat{\phi}}=e^g\sin\psi\sin\theta d\phi,
\end{eqnarray}
we obtain non-vanishing components of the spin connection
\begin{eqnarray}
{\omega^{\hat{t}}}_{\hat{r}}&=&f'e^{\hat{t}},\qquad {\omega^{\hat{\psi}}}_{\hat{r}}=g'e^{\hat{\psi}},\qquad {\omega^{\hat{\theta}}}_{\hat{r}}=g'e^{\hat{\theta}},\qquad {\omega^{\hat{\phi}}}_{\hat{r}}=g'e^{\hat{\phi}},\nonumber \\
{\omega^{\hat{\phi}}}_{\hat{\theta}}&=&e^{-g}\frac{\cot\theta}{\sin\psi}e^{\hat{\phi}},\qquad {\omega^{\hat{\phi}}}_{\hat{\psi}}=e^{-g}\cot\psi e^{\hat{\phi}},\qquad {\omega^{\hat{\theta}}}_{\hat{\psi}}=e^{-g}\cot{\psi}e^{\hat{\theta}}\, .
\end{eqnarray}
\indent We then turn on gauge fields corresponding to the $U(1)\times SU(2)_{\textrm{diag}}\subset U(1)\times SU(2)\times SU(2)$ symmetry and consider scalar fields that are singlet under $U(1)\times SU(2)_{\textrm{diag}}$. Using the coset representative \eqref{U1_SU2d_coset}, we find components of the composite connection that involve the gauge fields
\begin{equation}
{Q_i}^j=-\frac{i}{2}g_1A^0{(\sigma_2\otimes \sigma_3)_i}^j+\frac{i}{2}g_2\left[A^3{(\sigma_2\otimes \mathbb{I}_2)_i}^j-A^4{(\sigma_3\otimes \sigma_1)_i}^j+A^5{(\sigma_1\otimes \sigma_1)_i}^j\right].\label{composite_S3_compact}
\end{equation}
The components of the spin connection on $S^3$ that need to be cancelled are ${\omega^{\hat{\phi}}}_{\hat{\theta}}$, ${\omega^{\hat{\phi}}}_{\hat{\psi}}$ and ${\omega^{\hat{\theta}}}_{\hat{\psi}}$. To impose the twist, we set $A^0=0$ and take the $SU(2)_{\textrm{diag}}$ gauge fields to be
\begin{equation}
A^3=a_3\cos\psi d\theta,\qquad A^4=a_4\cos\theta d\phi,\qquad A^5=a_5\cos\psi\sin\theta d\phi\label{S3_gauge_fields}
\end{equation}
together with $A^{3+m}=\frac{g_2}{g_3}A^m$ for $m=3,4,5$. 
\\
\indent By considering the covariant derivative of $\epsilon_i$ along $\theta$ and $\phi$ directions, we find that the twist is achieved by imposing the following conditions 
\begin{equation}
g_2a_3=g_2a_4=g_2a_5=1\label{S3_twist}
\end{equation}
and projectors 
\begin{equation}
i\gamma_{\hat{\theta}\hat{\psi}}\epsilon_i={(\sigma_2\otimes \mathbb{I}_2)_i}^j\epsilon_j,\quad i\gamma_{\hat{\theta}\hat{\phi}}\epsilon_i={(\sigma_3\otimes \sigma_1)_i}^j\epsilon_j,\quad i\gamma_{\hat{\phi}\hat{\psi}}\epsilon_i={(\sigma_1\otimes \sigma_1)_i}^j\epsilon_j\, . \label{S3_projector}
\end{equation}
Note that the last projector is not independent of the first two. Therefore, the $AdS_2$ solutions preserve four supercharges of the original supersymmetry. Condition \eqref{S3_twist} also implies $a_3=a_4=a_5$. We will then set $a_3=a_4=a_5=a$ from now on. Using the definition \eqref{covariant_field_strength}, we find the gauge covariant field strengths
\begin{equation}
\mc{H}^3=-ae^{-2g}e^{\hat{\psi}}\wedge e^{\hat{\theta}},\quad \mc{H}^4=-ae^{-2g}e^{\hat{\theta}}\wedge e^{\hat{\phi}},\quad \mc{H}^5=-ae^{-2g}e^{\hat{\psi}}\wedge e^{\hat{\phi}}
\end{equation}
and $\mc{H}^{3+m}=\frac{g_2}{g_3}\mc{H}^m$ for $m=3,4,5$. 
\\
\indent For $\Sigma_3=H^3$, we use the metric ansatz
\begin{equation}
ds^2=-e^{2f}dt^2+dr^2+\frac{e^{2g}}{y^2}(dx^2+dy^2+dz^2)\label{metric_H3}
\end{equation}
with non-vanishing components of the spin connection
\begin{eqnarray}
{\omega^{\hat{x}}}_{\hat{r}}&=&g'e^{\hat{x}},\qquad {\omega^{\hat{y}}}_{\hat{r}}=g'e^{\hat{y}},\qquad {\omega^{\hat{z}}}_{\hat{r}}=g'e^{\hat{z}},\nonumber \\
{\omega^{\hat{x}}}_{\hat{y}}&=&-e^{-g}e^{\hat{x}},\qquad {\omega^{\hat{z}}}_{\hat{y}}=-e^{-g}e^{\hat{z}},\qquad {\omega^{\hat{t}}}_{\hat{r}}=f'e^{\hat{t}} 
\end{eqnarray}
where various components of the vielbein are given by
\begin{eqnarray}
e^{\hat{t}}&=&e^fdt,\qquad e^{\hat{r}}=dr,\qquad e^{\hat{x}}=\frac{e^g}{y}dx,\nonumber \\
e^{\hat{y}}&=&\frac{e^g}{y}dy,\qquad e^{\hat{z}}=\frac{e^g}{y}dz\, .
\end{eqnarray}
Since there are only two components, ${\omega^{\hat{x}}}_{\hat{y}}$ and ${\omega^{\hat{z}}}_{\hat{y}}$, of the spin connection to be cancelled in the twisting process, we turn on the following $SU(2)$ gauge fields
\begin{equation}
A^3=\frac{a}{y}dx,\qquad A^4=0,\qquad A^5=\frac{\tilde{a}}{y}dz \label{H3_gauge_fields}
\end{equation}
and $A^{m+3}=\frac{g_2}{g_3}A^m$, for $m=3,4,5$.
\\
\indent Repeating the same analysis as in the $S^3$ case, we find the twist conditions
\begin{equation}
g_2a=g_2\tilde{a}=1\label{twist_H3}
\end{equation}
and projectors
\begin{equation}
i\gamma_{\hat{y}\hat{x}}\epsilon_i={(\sigma_2\otimes \mathbb{I}_2)_i}^j\epsilon_j,\quad i\gamma_{\hat{y}\hat{z}}\epsilon_i={(\sigma_1\otimes \sigma_1)_i}^j\epsilon_j,\quad i\gamma_{\hat{x}\hat{z}}\epsilon_i={(\sigma_3\otimes \sigma_1)_i}^j\epsilon_j.\label{H3_compact_project}
\end{equation}
 The last projector is not needed for the twist with $A^4=0$. In addition, it follows from the first two projectors as in the $S^3$ case. The twist condition \eqref{twist_H3} again implies that $\tilde{a}=a$, and the covariant field strengths in this case are given by
\begin{equation}
\mc{H}^3=ae^{-2g}e^{\hat{x}}\wedge e^{\hat{y}},\qquad \mc{H}^4=ae^{-2g}e^{\hat{z}}\wedge e^{\hat{x}},\qquad \mc{H}^5=ae^{-2g}e^{\hat{z}}\wedge e^{\hat{y}}
\end{equation}
and $\mc{H}^{m+3}=\frac{g_2}{g_3}\mc{H}^m$, for $m=3,4,5$. Note that although $A^4=0$, we have non-vanishing $\mc{H}^4$ due to the non-abelian nature of $SU(2)$ field strengths. 
\\
\indent With all these ingredients, the following BPS equations are straightforwardly obtained
\begin{eqnarray}
\phi'&=&\frac{1}{8g_3}\Sigma^{-1}e^{-3\phi-2g}[g_2-g_3+e^{2\phi}(g_2+g_3)]\left[g_3e^{2g}(e^{4\phi}-1)+4\kappa ae^{2\phi}\Sigma^2\right],\quad\label{eq1_S3_compact}\\
\Sigma'&=&-\frac{1}{3}\left[g_2\cosh^3\phi+g_3\sinh^3\phi+\sqrt{2}g_1\Sigma^3\right]\nonumber \\
& &+\frac{\kappa}{g_3}ae^{-2g}\Sigma^2(g_3\cosh\phi+g_2\sinh\phi),\label{eq2_S3_compact}\\
g'&=&-\frac{1}{3}\Sigma^{-1}(g_2\cosh^3\phi+g_3\sinh^3\phi)+\frac{1}{3}g_1\Sigma^2\nonumber \\
& &-\frac{\kappa}{g_3}ae^{-2g}\Sigma (g_3\cosh\phi+g_2\sinh\phi),\label{eq3_S3_compact}\\
f'&=&-\frac{1}{3}\Sigma^{-1}(g_2\cosh^3\phi+g_3\sinh^3\phi)+\frac{1}{3}g_1\Sigma^2\nonumber \\
& &+\frac{\kappa}{g_3}ae^{-2g}\Sigma (g_3\cosh\phi+g_2\sinh\phi).\label{eq4_S3_compact}
\end{eqnarray}
As in the $AdS_3$ solutions, $\kappa=1$ and $\kappa=-1$ corresponds to $\Sigma_3=S^3$ and $\Sigma_3=H^3$, respectively.
\\
\indent It turns out that only $\kappa=-1$ leads to an $AdS_2$ solution given by
\begin{eqnarray}
\phi&=&\frac{1}{2}\ln\left[\frac{g_3-g_2}{g_3+g_2}\right],\qquad \Sigma=-\left[\frac{2\sqrt{2}g_2g_3}{g_1\sqrt{g_3^2-g_2^2}}\right]^{\frac{1}{3}},\nonumber \\
g&=&\frac{1}{2}\ln\left[\frac{2a(g_3^2-g_2^2)^{\frac{2}{3}}}{g_1^{\frac{2}{3}}g_2^{\frac{1}{3}}g_3^{\frac{4}{3}}}\right],\qquad L_{AdS_2}=\frac{(g_3^2-g_2^2)^{\frac{1}{3}}}{\sqrt{2}g_1^{\frac{1}{3}}g_2^{\frac{2}{3}}g_3^{\frac{2}{3}}}\, .\label{AdS2_compact}
\end{eqnarray} 
This solution preserves $N=4$ supersymmetry in two dimensions and $U(1)\times SU(2)_{\textrm{diag}}$ symmetry. As $r\rightarrow \infty$, $f\sim g\sim r$, solutions to the above BPS equations are locally asymptotic to either of the $N=4$ $AdS_5$ vacua in \eqref{AdS5_1_compact} and \eqref{AdS5_2_compact}. RG flow solutions interpolating between these $AdS_5$ vacua and the $AdS_2\times H^3$ solution in \eqref{AdS2_compact} are shown in figure \ref{fig7} and \ref{fig8}. In particular, the flow in figure \ref{fig8} connects three critical points similar to the solution given in the previous section.

\indent We end this section by a comment on the possibility of turning on the twist from $A^0$ along with those from the $SU(2)_{\textrm{diag}}$ gauge fields. As in the previous section, if we impose an additional projector
\begin{equation}
{(\mathbb{I}_2\otimes \sigma_3)_i}^j\epsilon_j=-\epsilon_i,
\end{equation}      
the projection matrix of the $A^0$ term in the composite connection \eqref{composite_S3_compact} will be proportional to that of $A^3$. We will consider the $S^3$ case for concreteness and take the ansatz for $A^0$ to be
\begin{equation}
A^0=a_0\cos\psi d\theta
\end{equation}
and proceed as in the $A^0=0$ case. This results in the projectors given in \eqref{S3_projector} and the twist conditions
\begin{equation}
g_2a_4=g_2a_5=1\qquad \textrm{and}\qquad g_1a_0+g_2a_3=1\, .\label{a0_twist_S3}
\end{equation}
We can see that at this stage the parameter $a_3$ needs not be equal to $a_4$ and $a_5$. However, consistency of the BPS equations from $\delta\lambda^a_i$ conditions require $a_3=a_4=a_5$ and hence $a_0=0$ by the conditions in \eqref{a0_twist_S3}. This is because $A^0$ does not appear in $\delta \lambda^a_i$ variation. The resulting BPS equations then reduce to those of the previous case with $A^0=0$. So, we conclude that the $A^0$ twist cannot be turned on along with the $SU(2)_{\textrm{diag}}$ twists.

\section{$U(1)\times SO(3,1)$ gauge group}\label{U1_SO3_1_gauge_group}
For non-compact $U(1)\times SO(3,1)$ gauge group, components of the embedding tensor are given by
\begin{eqnarray}
\xi^{MN}&=&g_1(\delta^M_2\delta^N_1-\delta^M_1\delta^N_2),\\ 
f_{345} &=&f_{378} = -f_{468} = -f_{567} = -g_2\, .\label{SO3_1_embedding}
\end{eqnarray} 
This gauge group has already been studied in \cite{5D_N4_flow}. The scalar potential admits one supersymmetric $N=4$ $AdS_5$ vauum at which all scalars from vector multiplets vanish and $\Sigma=1$ after choosing $g_2=-\sqrt{2}g_1$. At the vacuum, the gauge group is broken down to its maximal compact subgroup $U(1)\times SO(3)$. A holographic RG flow from this critical point to a non-conformal field theory in the IR and a flow to $AdS_3\times H^2$ vacuum preserving eight supercharges have also been studied in \cite{5D_N4_flow}. In this case, $AdS_3\times S^2$ solutions do not exist.
\\
\indent In this section, we will study $AdS_3\times \Sigma_2$ and $AdS_2\times \Sigma_3$ solutions preserving four supercharges. The analysis is closely parallel to that performed in the previous section, so we will give less detail in order to avoid repetition. 

\subsection{Supersymmetric black strings}
\begin{figure}
         \centering
         \begin{subfigure}[b]{0.3\textwidth}
                 \includegraphics[width=\textwidth]{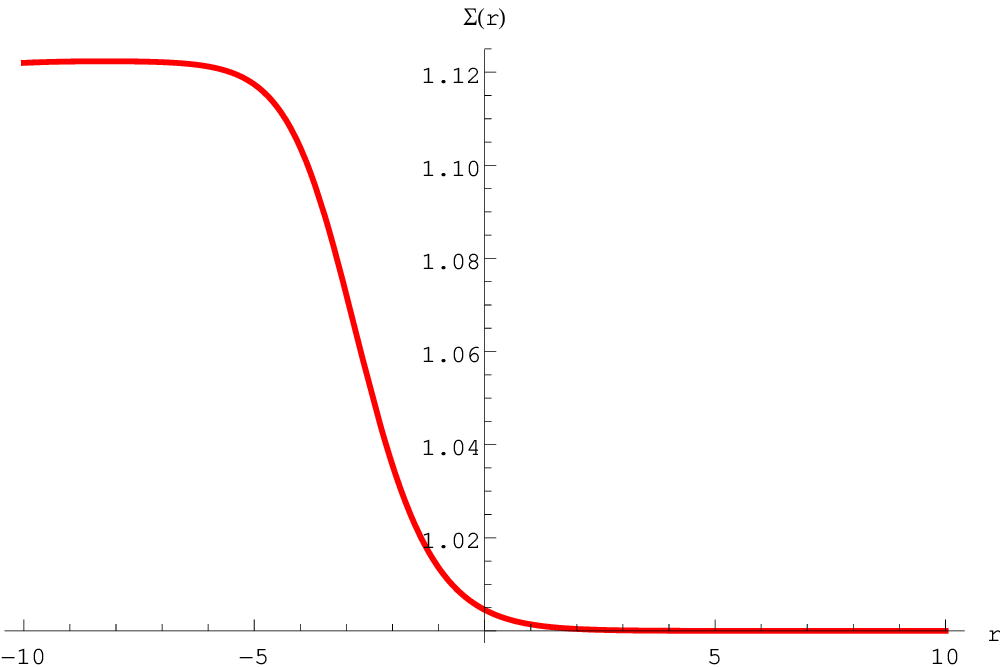}
                 \caption{Solution for $\Sigma$}
         \end{subfigure}\,\, 
\begin{subfigure}[b]{0.3\textwidth}
                 \includegraphics[width=\textwidth]{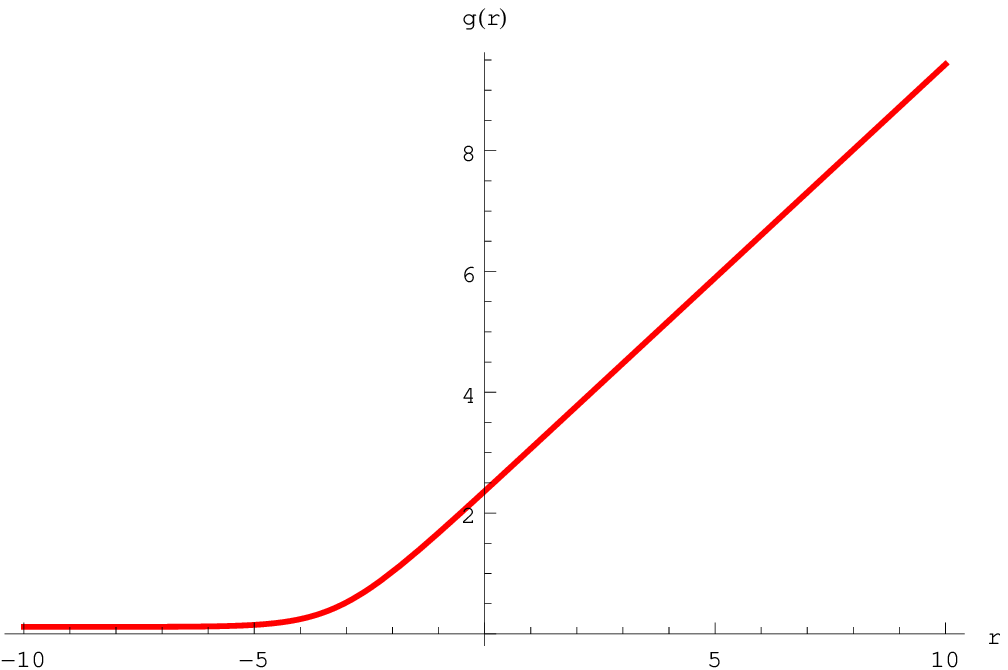}
                 \caption{Solution for $g$}
         \end{subfigure}\,\,
         ~ 
         \begin{subfigure}[b]{0.3\textwidth}
                 \includegraphics[width=\textwidth]{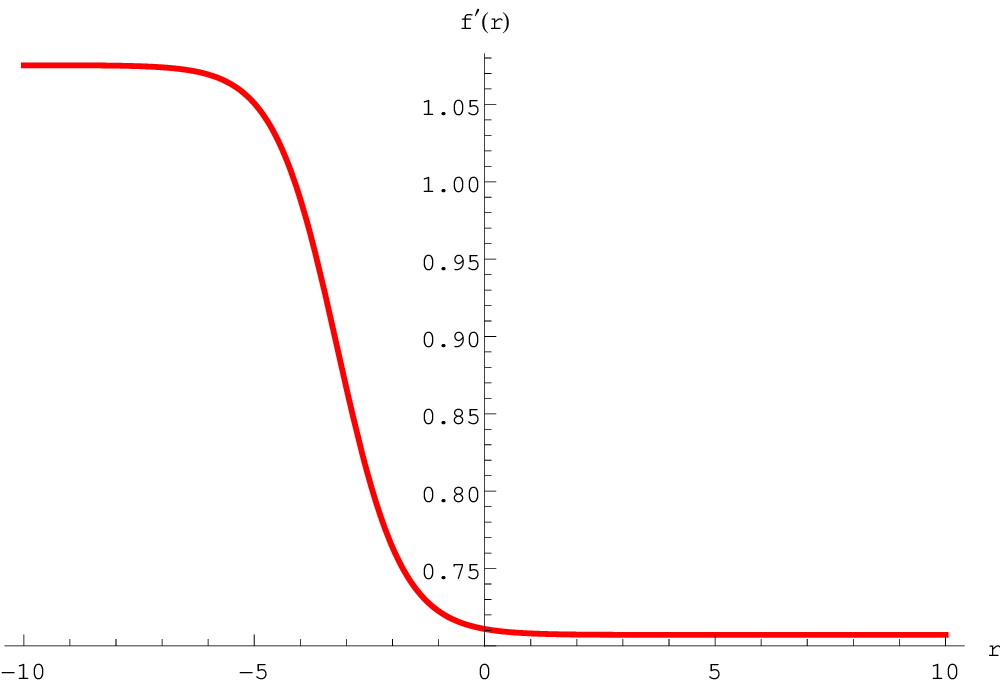}
                 \caption{Solution for $f'$}
         \end{subfigure}
         \caption{An RG flow solution from supersymmetric $AdS_5$ with $U(1)\times SO(3)$ symmetry to $AdS_3\times S^2$ geometry in the IR for $U(1)\times SO(3,1)$ gauge group and $g_1 = 1$, $a_5 = 1$.}\label{fig9}
 \end{figure}

We will use the same metric ansatz as in equations \eqref{metric_ansatz_AdS3S2} and \eqref{metric_ansatz_AdS3H2} and consider the twist from $U(1)\times U(1)$ gauge fields. The second $U(1)$ is a subgroup of the $SO(3)\subset SO(3,1)$. There are in total five scalars that are singlet under this $U(1)\times U(1)$, but as in the compact $U(1)\times SU(2)\times SU(2)$ gauge group, these can be truncated to three singlets corresponding to the following $SO(5,5)$ non-compact generators   
\begin{eqnarray}
\tilde{Y}_1 = Y_{31} + Y_{42}, \qquad
\tilde{Y}_2 = Y_{32}-Y_{41}, \qquad \tilde{Y}_3 = Y_{53}\, .
\end{eqnarray}
With the embedding tensor \eqref{SO3_1_embedding}, the compact $SO(3)$ symmetry is generated by $X_3$, $X_4$ and $X_5$ generators.
\\
\indent Using the coset representative of the form
\begin{eqnarray}
L = e^{\phi_1 \tilde{Y}_1}e^{\phi_2 \tilde{Y}_2}e^{\phi_3 \tilde{Y}_3},
\end{eqnarray}
we can repeat all the analysis of the previous section by using the ansatz for the gauge fields
\begin{equation}
A^0= a_0\cos\theta d\phi\qquad  \textrm{and}\qquad  A^5= a_5\cos\theta d\phi,
\end{equation}
for $\Sigma_2=S^2$ and  
\begin{equation}
A^0= a_0\cosh\theta d\phi\qquad  \textrm{and}\qquad  A^5= a_5\cosh\theta d\phi,
\end{equation}
for $\Sigma_2=H^2$. The result is similar to the compact case with the projectors \eqref{extra_projector1} and \eqref{theta_phi_projector1} and the twist condition \eqref{twist_con1}.
\\
\indent Using the $\gamma_r$ projection \eqref{gamma_r_projector}, the BPS equations in this case read
\begin{eqnarray}
f'&=& -\frac{1}{24 \Sigma^2}e^{-2 \phi_1-\phi_2-2 (\phi_3+g)} \left[e^{2 g} \left[1-e^{4 \phi_1}-e^{2 \phi_2}+e^{4 \phi_1+2 \phi_2}+e^{4 \phi_3}+4 e^{2 (\phi_1+\phi_3)}\right.\right.\nonumber\\
&&\left.\left.-\,\,e^{4 (\phi_1+\phi_3)}+4 e^{2 (\phi_1+\phi_2+\phi_3)}-e^{2 \phi_2+4 \phi_3}+e^{4 \phi_1+2 \phi_2+4 \phi_3}\right] g_2 \Sigma \right.\nonumber\\
&&-4 \sqrt{2}\kappa a_0 e^{2 \phi_1+\phi_2+2 \phi_3}-4\kappa a_5 e^{2 (\phi_1+\phi_3)} \left(1+e^{2 \phi_2}\right) \Sigma^3
\nonumber \\
& &\left.-4 \sqrt{2} e^{2 \phi_1+\phi_2+2 (\phi_3+g)} g_1 \Sigma^4\right],\\
g'&=& \frac{1}{24 \Sigma^2}e^{-2 \phi_1-\phi_2-2 (\phi_3+g)} \left[-e^{2 g} \left[1-e^{4 \phi_1}-e^{2 \phi_2}+e^{4 \phi_1+2 \phi_2}+e^{4 \phi_3}+4 e^{2 (\phi_1+\phi_3)}\right.\right.\nonumber\\
&&\left.\left.
-\,\,e^{4 (\phi_1+\phi_3)}+4 e^{2 (\phi_1+\phi_2+\phi_3)}-e^{2 \phi_2+4 \phi_3}+e^{4 \phi_1+2 \phi_2+4 \phi_3}\right] g_2 \Sigma  \right.\nonumber\\
&&
-8\kappa \sqrt{2} a_0 e^{2 \phi_1+\phi_2+2 \phi_3}-8\kappa a_5 e^{2 (\phi_1+\phi_3)} \left(1+e^{2 \phi_2}\right) \Sigma^3\nonumber \\
& &\left.+4 \sqrt{2} e^{2 \phi_1+\phi_2+2 (\phi_3+g)} g_1 \Sigma^4\right],\\
\Sigma '&=& \frac{1}{24 \Sigma}e^{-2 \phi_1-\phi_2-2 (\phi_3+g)} \left[-e^{2 g} \left(1-e^{4 \phi_1}-e^{2 \phi_2}+e^{4 \phi_1+2 \phi_2}+e^{4 \phi_3}+4 e^{2 (\phi_1+\phi_3)}\right.\right.\nonumber\\
&&\left.\left.-\,\,e^{4 (\phi_1+\phi_3)}+4 e^{2 (\phi_1+\phi_2+\phi_3)}-e^{2 \phi_2+4 \phi_3}+e^{4 \phi_1+2 \phi_2+4 \phi_3}\right) g_2 \Sigma 
\right.\nonumber\\
&&
-8\kappa \sqrt{2} a_0 e^{2 \phi_1+\phi_2+2 \phi_3}+4\kappa a_5 e^{2 (\phi_1+\phi_3)} \left(1+e^{2 \phi_2}\right) \Sigma^3\nonumber \\
& &\left.-8 \sqrt{2} e^{2 \phi_1+\phi_2+2 (\phi_3+g)} g_1 \Sigma^4\right],\\
\phi'_1&=& \frac{e^{-2 \phi_1-\phi_2+2 \phi_3} \left(1+e^{4 \phi_1}\right) \left(e^{2 \phi_2}-1\right) g_2}{2 \left(1+e^{4 \phi_3}\right) \Sigma},\\
\phi'_2&=& \frac{1}{8 \Sigma}e^{-2 \phi_1-\phi_2-2 (\phi_3+g)} \left[e^{2 g} \left(e^{4 \phi_1}-e^{2 \phi_2}+e^{4 \phi_1+2 \phi_2}-e^{4 \phi_3}-4 e^{2 (\phi_1+\phi_3)}\right.\right.\nonumber\\
&&\left.
+e^{4 (\phi_1+\phi_3)}-1+4 e^{2 (\phi_1+\phi_2+\phi_3)}-e^{2 \phi_2+4 \phi_3}+e^{4 \phi_1+2 \phi_2+4 \phi_3}\right) g_2\nonumber \\
& &
\left.+4\kappa a_5 e^{2 (\phi_1+\phi_3)} \left(e^{2 \phi_2}-1\right) \Sigma^2\right],\\
\phi'_3&=& \frac{e^{-2 \phi_1-\phi_2-2 \phi_3} \left(e^{4 \phi_1}-1\right) \left(e^{2 \phi_2}-1\right) \left(e^{4 \phi_3}-1\right) g_2}{8 \Sigma}\, .
\end{eqnarray}
This set of equations admits an $AdS_3$ solution given by
\begin{eqnarray}\label{eq:so31AdS3-1/4}
\phi_2&=& \phi_3= 0, \qquad 
 \Sigma= 
 \left(\frac{\sqrt{2}\kappa}{a_5 g_1}\right)^{\frac{1}{3}},  \nonumber\\
g&=& \frac{1}{3}\ln\left(\frac{\sqrt{2}a_5^2}{g_1} \right),\qquad L_{AdS_3} =\left(\frac{\sqrt{2} a_5^2}{g_1}\right)^{\frac{1}{3}} \frac{2}{\left(1 - \kappa a_5 g_2\right)}.
\end{eqnarray}
As in the compact case, $\Sigma_2$ can be either $S^2$ or $H^2$, depending on the values of $a_5$, $a_0$, $g_1$ and $g_2$ such that the twist condition \eqref{twist_con1} is satisfied. This is in contrast to the half-supersymmetric solution found in \cite{5D_N4_flow} for which only $\Sigma_2=H^2$ is possible.
\\
\indent To find a domain wall interpolating between the $AdS_5$ vacuum to this $AdS_3\times \Sigma_2$ solution, we further truncate the BPS equations by setting $\phi_i=0$ for $i=1,2,3$. The resulting equations are given by
\begin{eqnarray}\label{eq:ads5-3-1/4-so31}
f'&=& \frac{1}{6 \Sigma^2}e^{-2 g} \left( \sqrt{2}\kappa a_0-2 e^{2 g} g_2 \Sigma + 2\kappa a_5 \Sigma^3 -  \sqrt{2} e^{2 g} g_1 \Sigma^4\right),\\
g'&=&-\frac{1}{6\Sigma^2}e^{-2 g} \left(2 \sqrt{2}\kappa a_0+2 e^{2 g} g_2 \Sigma+4 \kappa a_5 \Sigma^3+ \sqrt{2} e^{2 g} g_1 \Sigma^4\right),\\
\Sigma '&=&-\frac{1}{3 \Sigma}e^{-2 g} \left(\sqrt{2}\kappa a_0+ e^{2 g} g_2 \Sigma-\kappa a_5 \Sigma^3+ \sqrt{2} e^{2 g} g_1 \Sigma^4\right).\end{eqnarray}
An example of numerical solutions is shown in figure \ref{fig9}.

\subsection{Supersymmetric black holes}
\begin{figure}
         \centering
         \begin{subfigure}[b]{0.3\textwidth}
                 \includegraphics[width=\textwidth]{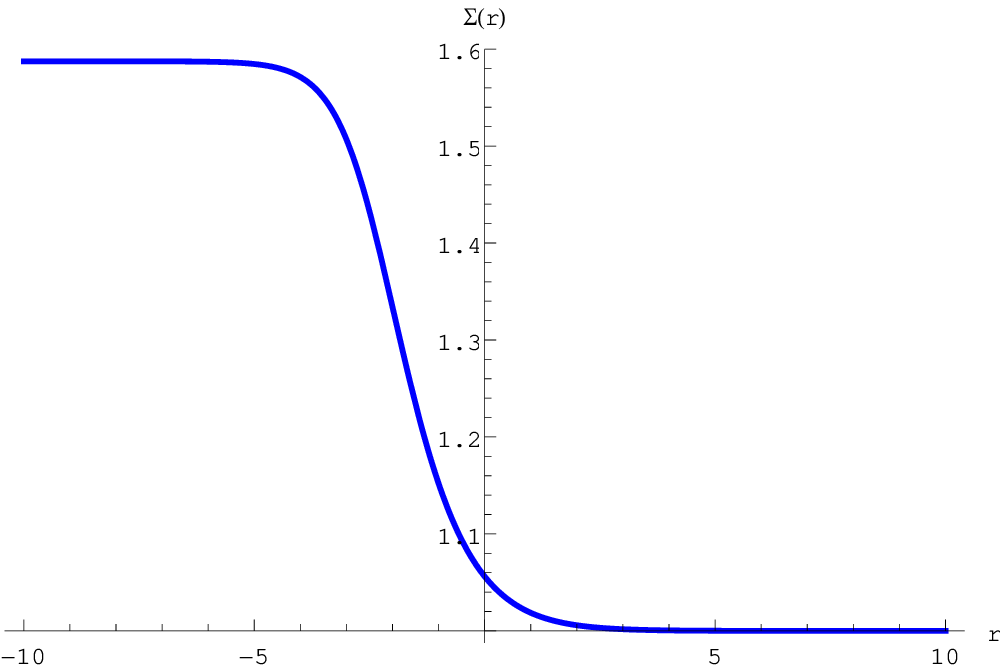}
                 \caption{Solution for $\Sigma$}
         \end{subfigure}\,\, 
\begin{subfigure}[b]{0.3\textwidth}
                 \includegraphics[width=\textwidth]{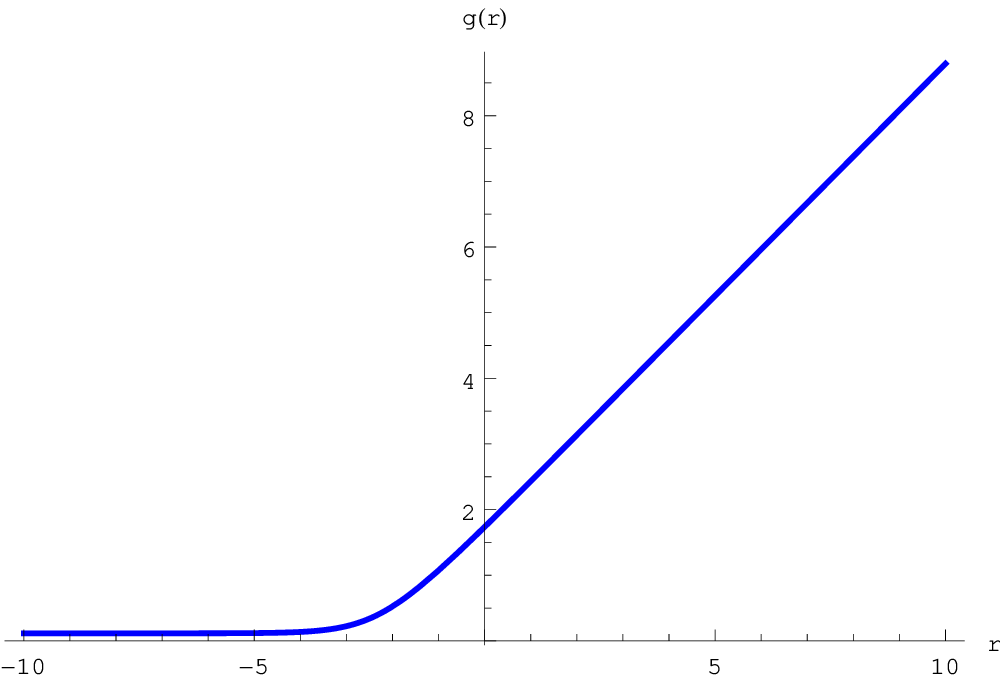}
                 \caption{Solution for $g$}
         \end{subfigure}\,\,
         ~ 
         \begin{subfigure}[b]{0.3\textwidth}
                 \includegraphics[width=\textwidth]{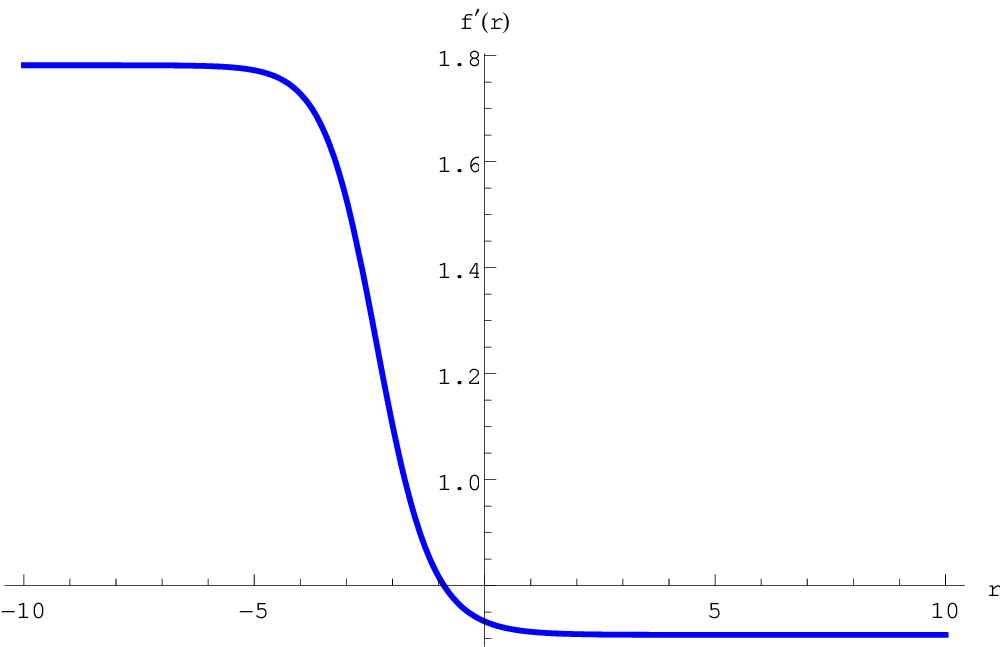}
                 \caption{Solution for $f'$}
         \end{subfigure}
         \caption{An RG flow solution from $AdS_5$ with $U(1)\times SO(3)$ symmetry to $AdS_2\times H^3$ geometry in the IR for $U(1)\times SO(3,1)$ gauge group and $g_1 = 1$.}\label{fig10}
 \end{figure}

We now consider $AdS_2\times \Sigma_3$ solutions within this non-compact gauge group. We will look for solutions with $U(1)\times SO(3)\subset U(1)\times SO(3,1)$ symmetry. There is one $U(1)\times SO(3)$ singlet from the $SO(5,5)/SO(5)\times SO(5)$ coset corresponding to the non-compact generator
\begin{eqnarray}
Y = Y_{31} +Y_{42} - Y_{53}\, .
\end{eqnarray}
The coset representative can be written as
\begin{eqnarray}
L = e^{\phi Y}\, .
\end{eqnarray}
Using the metric ansatz \eqref{metric_S3} and \eqref{metric_H3} together with the gauge fields \eqref{S3_gauge_fields} and \eqref{H3_gauge_fields}, we find that the twist can be implemented by using the projectors given in \eqref{S3_projector}. Furthermore, the twist condition also implies that $a_3=a_4=a_5=a$ with $g_2a=1$, and the twist from $A^0$ cannot be turned on. The $AdS_2\times \Sigma_3$ solutions preserve four supercharges.
\\
\indent Using the projector \eqref{gamma_r_projector}, we can derive the following BPS equations
\begin{eqnarray}
f'&=& \frac{1}{12 \Sigma}\left[e^{-3 \phi} \left(1-3 e^{2 \phi}-3 e^{4 \phi}+e^{6 \phi}\right) g_2+6\kappa a e^{-\phi-2 g} \left(1+e^{2 \phi}\right) \Sigma^2\right.\nonumber \\
& &\left.+2 \sqrt{2} g_1 \Sigma^3\right],
\\
g'&=& \frac{1}{12 \Sigma}\left[e^{-3 \phi} \left(1-3 e^{2 \phi}-3 e^{4 \phi}+e^{6 \phi}\right) g_2-6\kappa a e^{-\phi-2 g} \left(1+e^{2 \phi}\right) \Sigma^2\right.\nonumber \\
& &\left.+2 \sqrt{2} g_1 \Sigma^3\right],
\\
\Sigma'&=& \frac{1}{12} e^{-3 \phi-2 g} \left[e^{2 g} \left(1-3 e^{2 \phi}-3 e^{4 \phi}+e^{6 \phi}\right) g_2+6\kappa a e^{2 \phi} \left(1+e^{2 \phi}\right) \Sigma^2\right.\nonumber \\
& &\left.-4 \sqrt{2} e^{3 \phi+2 g} g_1 \Sigma^3\right],
\\
\phi'&=& -\frac{1}{4 \Sigma}e^{-3 \phi-2 g} \left(e^{2 \phi}-1\right) \left(e^{2 g} \left(1+e^{4 \phi}\right) g_2-2\kappa a e^{2 \phi} \Sigma^2\right).
\end{eqnarray}
These equations admit one $AdS_2\times H^3$ solution given by
\begin{eqnarray}\label{eq:so31AdS2}
\phi  &=&0, \qquad \Sigma = -\sqrt{2}\left(\frac{g_2}{g_1}\right)^{\frac{1}{3}}
\nonumber\\
g &=& -\frac{1}{2}\ln\left[\frac{\left(\frac{}{}g_1^{2}g_2\right)^{\frac{1}{3}}}{2a}\right],\qquad L_{AdS_2} =\frac{1}{\sqrt{2}\left(g_1 g^2_2\right)^{\frac{1}{3}}}\label{AdS2_SO3_1}
\end{eqnarray}
while $AdS_2\times S^3$ solutions do not exist.
\\
\indent By setting $\phi=0$, we find a numerical solution to the above BPS equations as shown in figure \ref{fig10}.
   
\section{$U(1)\times SL(3,\mathbb{R})$ gauge group}\label{U1_SL3_R_gauge_group}
In this section, we consider non-compact $U(1)\times SL(3,\mathbb{R})$ gauge group. This has not been studied in \cite{5D_N4_flow}, so we will give more detail about the construction of this gauged supergravity and possible supersymmetric $AdS_5$ vacua.
\\
\indent Components of the embedding tensor for this gauge group are given by 
\begin{eqnarray}\label{eq:sl3eb}
\xi^{MN} &=&g_1(\delta^M_2\delta^N_1 -\delta^M_1\delta^N_2),\label{U1_SL3}\\
f_{345} &=& f_{389} =f_{468} =f_{497}= f_{569}=f_{578} =-g_2,\nonumber\\
f_{367} &=& 2g_2, \qquad f_{4,9,10} =f_{5,8,10}= \sqrt{3}g_2\, .\label{SL3_embedding}
\end{eqnarray}  
${f_{MN}}^P$ can be extracted from $SL(3,\mathbb{R})$ generators $(\frac{i\lambda_2}{2},\frac{i\lambda_5}{2},\frac{i\lambda_7}{2},\frac{\lambda_1}{2},\frac{\lambda_3}{2},\frac{\lambda_4}{2},\frac{\lambda_6}{2},\frac{\lambda_8}{2})$ with $\lambda_i$, $i=1,2,\ldots, 8$ being the usual Gell-Mann matrices. The compact $SO(3)\subset SL(3,\mathbb{R})$ symmetry is generated by $X_3$, $X_4$ and $X_5$.  

\subsection{Supersymmetric $AdS_5$ vacuum}
The $SL(3,\mathbb{R})$ factor is embedded in $SO(3,5)\subset{SO(5,5)}$ such that its adjoint representation is identified with the fundamental representation of $SO(3,5)$. The $SO(3)\subset SL(3,\mathbb{R})$ is embedded in $SL(3,\mathbb{R})$ such that $\mathbf{3}\rightarrow \mathbf{3}$. Decomposing the adjoint representation of $SO(3,5)$ to $SL(3,\mathbb{R})$ and $SO(3)$, we find that the $25$ scalars transform under $SO(3)\subset SL(3,\mathbb{R})$ as
\begin{equation}
2(\mathbf{1}\times \mathbf{5})+\mathbf{3}\times \mathbf{5}=\mathbf{3}+3\times \mathbf{5}+\mathbf{7}\, .
\end{equation}
Unlike the $U(1)\times SO(3,1)$ gauge group, there is no singlet under the compact $SO(3)$ symmetry. Taking into account the embedding of the $U(1)$ factor in the gauge group as described in \eqref{U1_SL3}, we find the transformation of the scalars under $U(1)\times SO(3)$ 
\begin{equation}
\mathbf{3}_0+\mathbf{5}_0+\mathbf{7}_0+\mathbf{5}_2+\mathbf{5}_{-2}
\end{equation}
with the subscript denoting the $U(1)$ charges.
\\
\indent It can be readily verified by studying the corresponding scalar potential or recalling the result of \cite{AdS5_N4_Jan} that this $U(1)\times SL(3,\mathbb{R})$ gauge group admits a supersymmetric $N=4$ $AdS_5$ vacuum at which all scalars from vector multiplets vanish with
\begin{equation}
\Sigma=1\qquad \textrm{and}\qquad V_0=-3g_1^2\, .
\end{equation}    
We have, as in other gauge groups, set $g_2=-\sqrt{2}g_1$ to bring this vacuum to the value of $\Sigma=1$. All scalar masses at this vacuum are given in table \ref{table1}. Massless scalars in $\mathbf{5}_0$ representation are Goldstone bosons corresponding to the symmetry breaking $SL(3,\mathbb{R})\rightarrow SO(3)$.    
\begin{table}[h]
\centering
\begin{tabular}{|c|c|c|}
  \hline
  Scalar field representations & $m^2L^2\phantom{\frac{1}{2}}$ & $\Delta$  \\ \hline
  $\mathbf{1}_0$ & $-4$ &  $2$  \\
  $\mathbf{3}_0$ & $32$ &  $8$  \\
  $\mathbf{5}_0$ & $0$ &  $4$  \\
   $\mathbf{7}_0$ & $12$ &  $6$  \\
    $\mathbf{5}_{-2}$ & $21$ &  $7$  \\
      $\mathbf{5}_2$ & $21$ &  $7$  \\
  \hline
\end{tabular}
\caption{Scalar masses at the $N=4$ supersymmetric $AdS_5$ critical point with $U(1)\times SO(3)$ symmetry and the corresponding dimensions of dual operators for the non-compact $U(1)\times SL(3,\mathbb R)$ gauge group. The scalars are organized into representations of $U(1)\times SO(3)$ with the singlet corresponding to the dilaton $\Sigma$.}\label{table1}
\end{table} 
      
\subsection{Supersymmetric black strings}           
\begin{figure}
         \centering
         \begin{subfigure}[b]{0.3\textwidth}
                 \includegraphics[width=\textwidth]{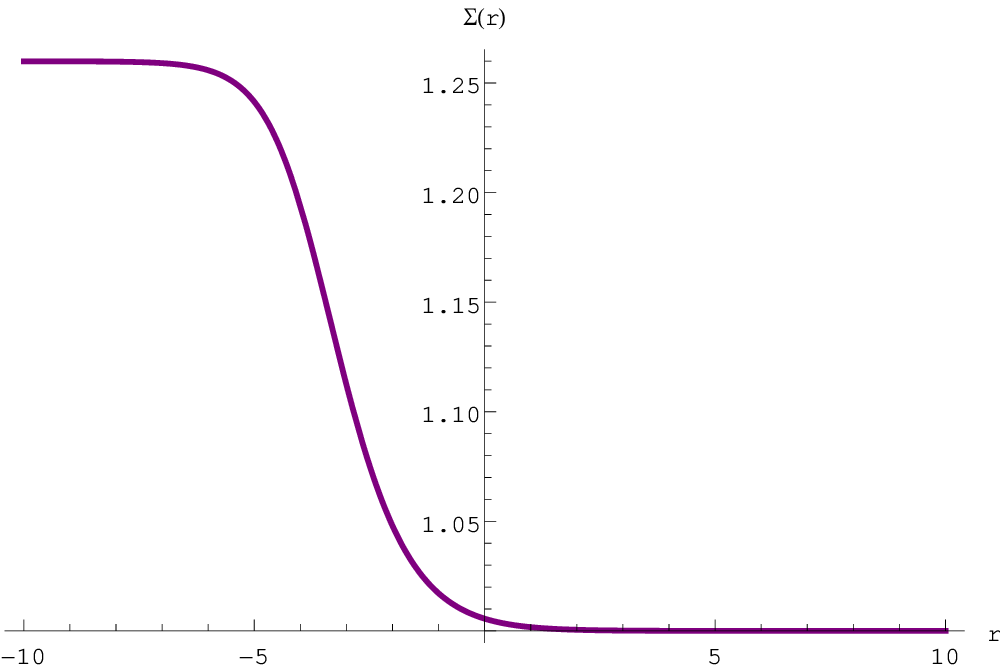}
                 \caption{Solution for $\Sigma$}
         \end{subfigure}\,\, 
\begin{subfigure}[b]{0.3\textwidth}
                 \includegraphics[width=\textwidth]{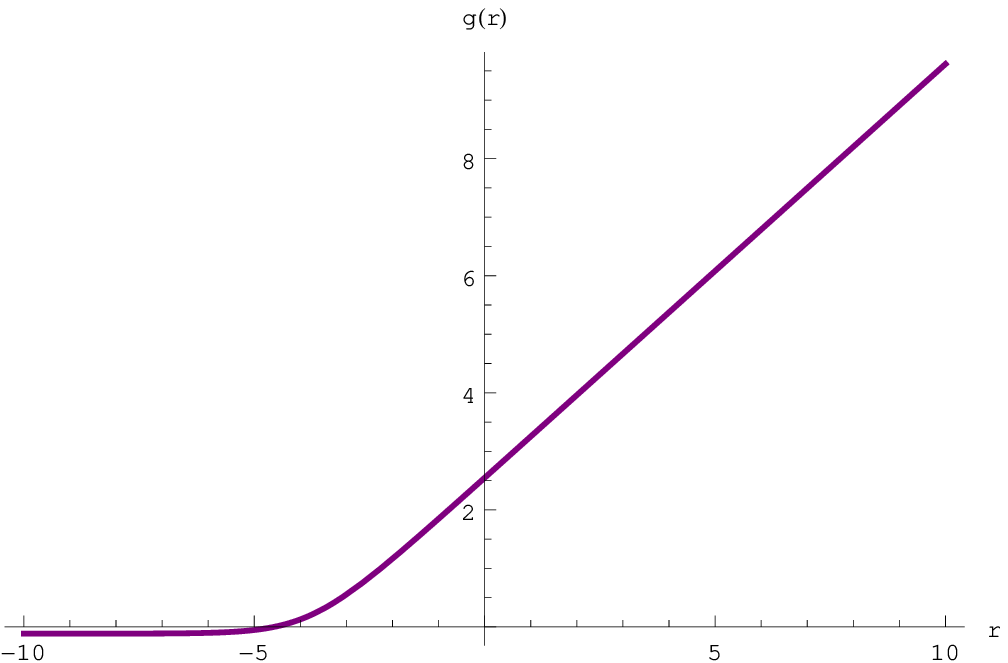}
                 \caption{Solution for $g$}
         \end{subfigure}\,\,
         ~ 
         \begin{subfigure}[b]{0.3\textwidth}
                 \includegraphics[width=\textwidth]{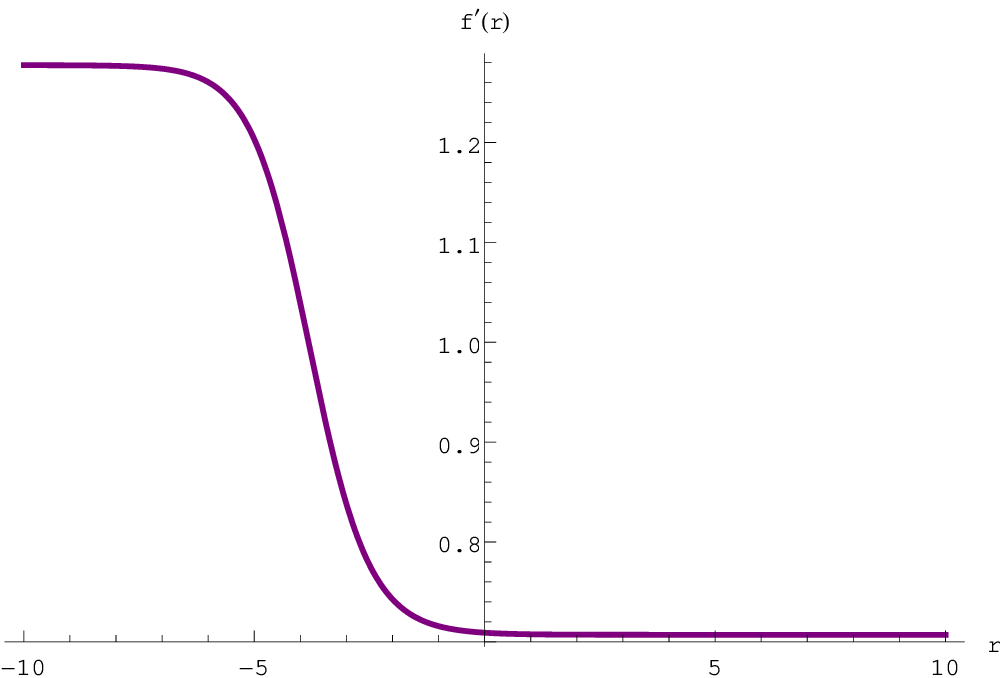}
                 \caption{Solution for $f'$}
         \end{subfigure}
         \caption{An RG flow solution from $AdS_5$ with $U(1)\times SO(3)$ symmetry to $N=4$ $AdS_3\times H^2$ geometry in the IR for $U(1)\times SL(3,\mathbb{R})$ gauge group and $g_1 = 1$.}\label{fig11}
 \end{figure}

We now consider $U(1)\times U(1)\subset U(1)\times SO(3)\subset U(1)\times SL(3,\mathbb{R})$ invariant scalars. We will choose the $U(1)\subset SO(3)$ generator to be $X_5$. From the vector multiplets, there are three singlet scalars corresponding to the following non-compact generators
\begin{eqnarray}
\bar{Y}_1 = Y_{31}-Y_{44}, \qquad \bar{Y}_2 = Y_{41}+ Y_{34}, \qquad \bar{Y}_3 = \sqrt{3} Y_{52} - Y_{55}\, .
\end{eqnarray}
The coset representative can be written as
\begin{eqnarray}
L = e^{\phi_1 \bar{Y}_1} e^{\phi_2 \bar{Y}_2} e^{\phi_3 \bar{Y}_3}
\end{eqnarray}
which gives rise to the scalar potential
\begin{eqnarray}\label{eq:sl3VT5}
V&=&\frac{1}{16 \Sigma^2}e^{-4 (\phi_2+\phi_3)} g_2 \left[\left(3+6 e^{4 \phi_2}+3 e^{8 \phi_2}+3 e^{8 \phi_3}-32 e^{4 (\phi_2+\phi_3)}+3 e^{8 (\phi_2+\phi_3)}\right.\right. \nonumber \\
& &\left.+6 e^{4 \phi_2+8 \phi_3}\right) g_2-4 \sqrt{2} e^{2 (\phi_2+\phi_3)} \left(\sqrt{3}-2 e^{2 \phi_2}-\sqrt{3} e^{4 \phi_2}-\sqrt{3} e^{4 \phi_3}\right.\nonumber \\
& &\left.\left.+\sqrt{3} e^{4 (\phi_2+\phi_3)}-2 e^{2 \phi_2+4 \phi_3}\right) g_1 \Sigma^3\right].
\end{eqnarray}
Notice that $V$ doesn't depend on $\phi_1$, consistent with the fact that $\phi_1$ is part of the Goldstone bosons in $\mathbf 5_0$ representation. It can be verified that this potential admits only one supersymmetric $AdS_5$ critical point at $\phi_1=\phi_2=\phi_3=0$ and $\Sigma=1$ for $g_2=-\sqrt{2}g_1$.
\\
\indent We first consider $AdS_3\times \Sigma_2$ solutions preserving eight supercharges. We will omit some detail since the same analysis has been carried out in \cite{5D_N4_flow}. By turning on gauge fields $A^0$ and $A^5$ along $\Sigma_2$ and performing the twist in equation \eqref{Covariant_Sigma2} by imposing only one projector 
\begin{equation}
i\gamma_{\hat{\theta}\hat{\phi}}\epsilon_i=a_0g_1{(\sigma_2\otimes\sigma_3)_i}^j\epsilon_j-a_5g_2{(\sigma_1\otimes \sigma_1)_i}^j\epsilon_j,
\end{equation}
we find that consistency of this projection condition, namely $(i\gamma_{\hat{\phi}\hat{\theta}})^2=\mathbb{I}_4$, implies $a_0a_5=0$, see \cite{5D_N4_flow} for more detail. Therefore, for half-supersymmetric solutions, the twists from $A^0$ and $A^5$ cannot be turned on simultaneously. Furthermore, as shown in \cite{5D_N4_flow}, see also a similar discussion in \cite{flow_acrossD_bobev}, the twist with $a_5=0$ does not lead to an $AdS_3$ fixed point. We will accordingly consider only the case of $a_0=0$ and $a_5\neq 0$ which leads to the twist condition $a_5g_2=1$ and the projector 
\begin{equation}
i\gamma_{\hat{\theta}\hat{\phi}}\epsilon_i=-{(\sigma_1\otimes \sigma_1)_i}^j\epsilon_j\, .
\end{equation}
The resulting BPS equations read
\begin{eqnarray}
f'&=& \frac{1}{12 \Sigma}e^{-2 (\phi_2+\phi_3+g)} \left[e^{2 g} \left(\sqrt{3}-2 e^{2 \phi_2}-\sqrt{3} e^{4 \phi_2}-\sqrt{3} e^{4 \phi_3}+\sqrt{3} e^{4 (\phi_2+\phi_3)}\right.\right.\nonumber\\
& &\left.\left.-2 e^{2 \phi_2+4 \phi_3}\right) g_2+2\kappa a_5 e^{2 \phi_2} \left(1+e^{4 \phi_3}\right) \Sigma^2+2 \sqrt{2} e^{2 (\phi_2+\phi_3+g)} g_1 \Sigma^3\right],\\
g'&=& \frac{1}{12 \Sigma}e^{-2 (\phi_2+\phi_3+g)} \left[e^{2 g} \left(\sqrt{3}-2 e^{2 \phi_2}-\sqrt{3} e^{4 \phi_2}-\sqrt{3} e^{4 \phi_3}+\sqrt{3} e^{4 (\phi_2+\phi_3)} \right.\right.\nonumber\\
& &\left.\left.-2 e^{2 \phi_2+4 \phi_3}\right) g_2-4\kappa a_5 e^{2 \phi_2} \left(1+e^{4 \phi_3}\right) \Sigma^2+2 \sqrt{2} e^{2 (\phi_2+\phi_3+g)} g_1 \Sigma^3\right],\\
\Sigma '&=& \frac{1}{12} e^{-2 (\phi_2+\phi_3+g)} \left[e^{2 g} \left(\sqrt{3}-2 e^{2 \phi_2}-\sqrt{3} e^{4 \phi_2}-\sqrt{3} e^{4 \phi_3}+\sqrt{3} e^{4 (\phi_2+\phi_3)}\right.\right.\nonumber\\
& &\left.\left.-2 e^{2 \phi_2+4 \phi_3}\right) g_2+2\kappa a_5 e^{2 \phi_2} \left(1+e^{4 \phi_3}\right) \Sigma^2-4 \sqrt{2} e^{2 (\phi_2+\phi_3+g)} g_1 \Sigma^3\right],\\	
\phi'_1&=& 0\\
\phi'_2&=& -\frac{\sqrt{3} e^{-2 (\phi_2+\phi_3)} \left(1+e^{4 \phi_2}\right) \left(e^{4 \phi_3}-1\right) g_2}{4 \Sigma},\\
\phi'_3&=& -\frac{1}{8 \Sigma}e^{-2 (\phi_2+\phi_3+g)} \left[e^{2 g} \left(2 e^{2 \phi_2}-\sqrt{3}+\sqrt{3} e^{4 \phi_2}-\sqrt{3} e^{4 \phi_3}+\sqrt{3} e^{4 (\phi_2+\phi_3)}\right.\right.\nonumber\\
& &\left.\left.-2 e^{2 \phi_2+4 \phi_3}\right) g_2-2\kappa a_5 e^{2 \phi_2} \left(e^{4 \phi_3}-1\right) \Sigma^2\right].
\end{eqnarray}
The Killing spinors $\epsilon_i$ are subject to the projection conditions \eqref{gamma_r_projector} and
\begin{equation}\label{eq:bps-1/2}
i\gamma_{\hat\theta\hat\phi}\epsilon_i= -{(\sigma_1\otimes \sigma_1)_i}^j\epsilon_j\, .
\end{equation}
As in the $U(1)\times SO(3,1)$ gauge group studied in \cite{5D_N4_flow}, there is only one supersymmetric $AdS_3\times H^2$ critical point given by
\begin{eqnarray}\label{eq:sl3AdS3-1/2-T5}
\phi_1 &=& \phi_2 = \phi_3 = 0, \qquad \Sigma =-\left(\frac{\sqrt{2}\,g_2}{g_1}\right)^{\frac{1}{3}}, \nonumber\\
g&=&-\frac{1}{2}\ln\left[\frac{1}{a_5}\left(\frac{g_1^2 g_2}{2}\right)^{\frac{1}{3}}\right] \qquad L_{AdS_3} = \left(\frac{\sqrt{2}}{g_1g_2^2}\right)^{\frac{1}{3}}\, .
\end{eqnarray}                         
This solution is dual to a two-dimensional $N=(2,2)$ SCFT. By setting $\phi_1=\phi_2=\phi_3=0$, we find a domain wall interpolating between this critical point and the supersymmetric $AdS_5$ as shown in figure \ref{fig11}.                                
                    
\begin{figure}
         \centering
         \begin{subfigure}[b]{0.3\textwidth}
                 \includegraphics[width=\textwidth]{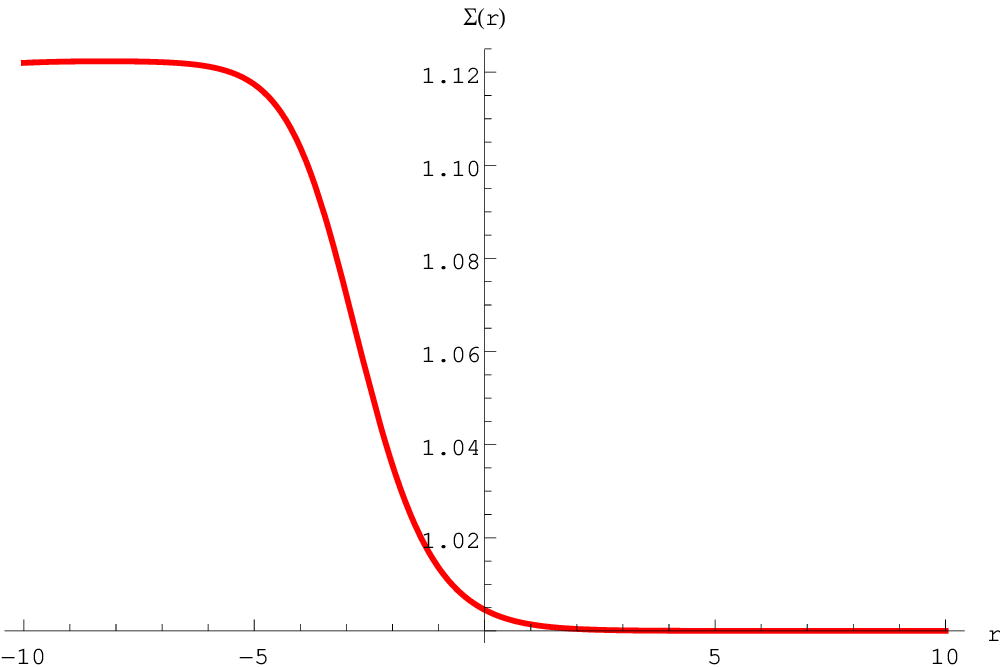}
                 \caption{Solution for $\Sigma$}
         \end{subfigure}\,\, 
\begin{subfigure}[b]{0.3\textwidth}
                 \includegraphics[width=\textwidth]{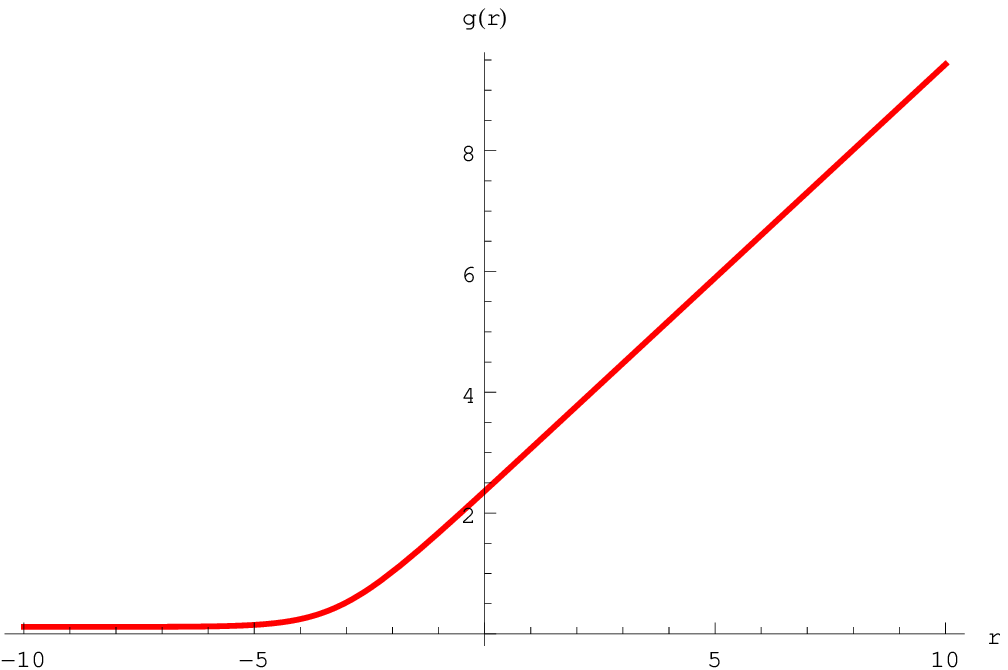}
                 \caption{Solution for $g$}
         \end{subfigure}\,\,
         ~ 
         \begin{subfigure}[b]{0.3\textwidth}
                 \includegraphics[width=\textwidth]{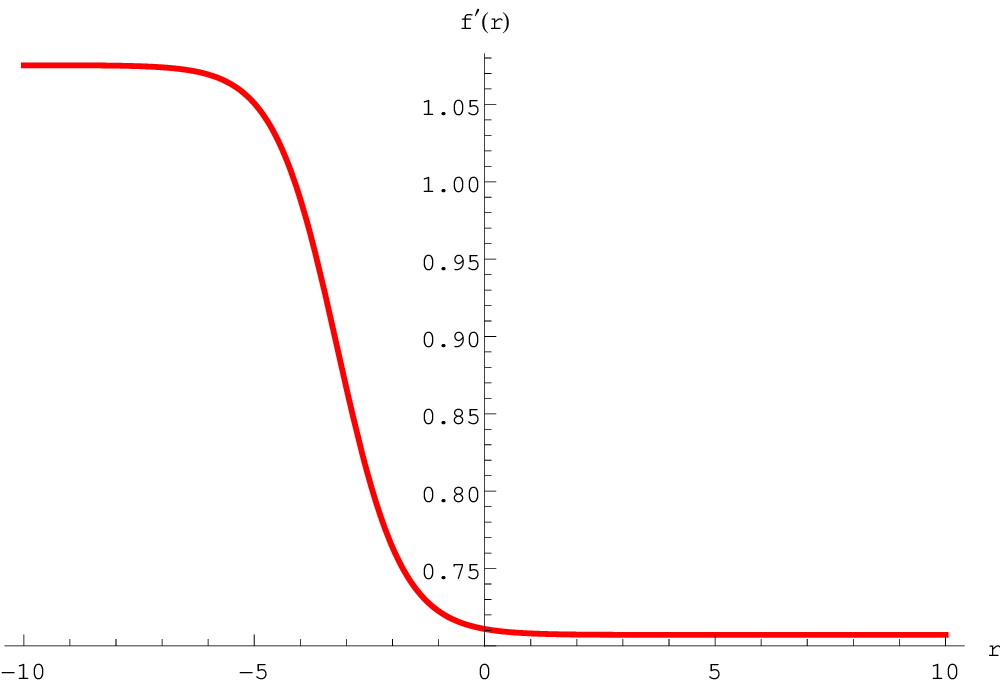}
                 \caption{Solution for $f'$}
         \end{subfigure}
         \caption{An RG flow solution from $AdS_5$ with $U(1)\times SO(3)$ symmetry to $N=2$ $AdS_3\times S^2$ geometry in the IR for $U(1)\times SL(3,\mathbb{R})$ gauge group and $g_1 = 1$, $a_5 = 1$.}\label{fig12}
 \end{figure}

\indent We now move to $AdS_3\times \Sigma_2$ solutions preserving four supercharges. The analysis follows the same line as in the previous two gauge groups, so we will be very brief in this section. By the same analysis as in the previous two gauge groups, we obtain the following BPS equations
\begin{eqnarray}
f'&=& \frac{1}{12 \Sigma^2}e^{-2 (\phi_2+\phi_3+g)} \left[2 \sqrt{2}\kappa a_0 e^{2 (\phi_2+\phi_3)}-e^{2 g} \left(\sqrt{3}-2 e^{2 \phi_2}
-\sqrt{3} e^{4 \phi_2}-\sqrt{3} e^{4 \phi_3}\right.\right.\nonumber\\&&\hspace{5mm}\left.\left.+\,\,\sqrt{3} e^{4 (\phi_2+\phi_3)}-2 e^{2 \phi_2+4 \phi_3}\right)g_2 \Sigma +\,\, 2\kappa a_5 e^{2 \phi_2} \left(1+e^{4 \phi_3}\right) \Sigma^3\right. \nonumber\\
& &\left.+2 \sqrt{2} e^{2 (\phi_2+\phi_3+g)}g_1 \Sigma^4\right],\\
g'&=& -\frac{1}{12 \Sigma^2}e^{-2 (\phi_2+\phi_3+g)} \left[4 \sqrt{2}\kappa a_0 e^{2 (\phi_2+\phi_3)}+e^{2 g} \left(\sqrt{3}-2 e^{2 \phi_2}-\sqrt{3} e^{4 \phi_2}-\sqrt{3} e^{4 \phi_3}\right.\right.\nonumber\\&&\hspace{5mm}\left.\left.+\sqrt{3} e^{4 (\phi_2+\phi_3)}-2 e^{2 \phi_2+4 \phi_3}\right)g_2 \Sigma + 4\kappa a_5 e^{2 \phi_2} \left(1+e^{4 \phi_3}\right) \Sigma^3\right. \nonumber\\
& &\left.-2 \sqrt{2} e^{2 (\phi_2+\phi_3+g)}g_1 \Sigma^4\right],\\
\Sigma '&=& \frac{1}{12 \Sigma}e^{-2 (\phi_2+\phi_3+g)} \left[-4 \sqrt{2}\kappa a_0 e^{2 (\phi_2+\phi_3)}+e^{2 g} \left(\sqrt{3}-2 e^{2 \phi_2}-\sqrt{3} e^{4 \phi_2}-\sqrt{3} e^{4 \phi_3}\right.\right.\nonumber\\&&\hspace{5mm}\left.\left.+\,\,\sqrt{3} e^{4 (\phi_2+\phi_3)}-2 e^{2 \phi_2+4 \phi_3}\right)g_2 \Sigma+2\kappa a_5 e^{2 \phi_2} \left(1+e^{4 \phi_3}\right) \Sigma^3\right. \nonumber\\
& &\left.-4 \sqrt{2} e^{2 (\phi_2+\phi_3+g)}g_1 \Sigma^4\right],
\end{eqnarray}
\begin{eqnarray}
\phi'_1&=& 0,\\
\phi'_2&=& -\frac{\sqrt{3} e^{-2 (\phi_2+\phi_3)} \left(1+e^{4 \phi_2}\right) \left(e^{4 \phi_3}-1\right)g_2}{4 \Sigma},\\
\phi'_3&=& -\frac{1}{8 \Sigma}e^{-2 (\phi_2+\phi_3+g)} \left[e^{2 g} \left(2 e^{2 \phi_2}-\sqrt{3}+\sqrt{3} e^{4 \phi_2}-\sqrt{3} e^{4 \phi_3}+\sqrt{3} e^{4 (\phi_2+\phi_3)}\right.\right. \nonumber \\
& &\left.\left.-2 e^{2 \phi_2+4 \phi_3}\right)g_2-2\kappa a_5 e^{2 \phi_2} \left(e^{4 \phi_3}-1\right) \Sigma^2\right].
\end{eqnarray}
These equations admit one supersymmetric $AdS_3\times \Sigma_2$ solution given by
\begin{eqnarray}\label{eq:sl3AdS3-1/4}
\phi_2&=& \phi_3= 0, \qquad  
\Sigma= \left(\frac{\sqrt{2}\kappa}{a_5 g_1}\right)^{\frac{1}{3}},  
\nonumber\\
g&=& 
\frac{1}{3}\ln\left(\frac{\sqrt{2} a_5^2}{g_1}\right),
 \qquad L_{AdS_3} =\left(\frac{\sqrt{2} a_5^2}{g_1}\right)^{\frac{1}{3}} \frac{2}{\left(1 -\kappa a_5 g_2\right)},
\end{eqnarray}
and a domain wall interpolating between this critical point and the supersymmetric $AdS_5$ is shown in figure \ref{fig12}. It should also be noted that this $AdS_3\times \Sigma_2$ solution is the same as in $U(1)\times SO(3,1)$ gauge group. 

\subsection{Supersymmetric black holes}
We end this section with an analysis of $AdS_2\times \Sigma_3$ solutions and domain walls connecting these solutions to the supersymmetric $AdS_5$. In order to preserve supersymmetry, $SO(3)\subset SL(3,\mathbb{R})$ gauge fields must be turned on. However, in the present case, there is no $SO(3)$ singlet scalar from the vector multiplets. After using the twist condition $g_2a=1$ and projectors in \eqref{S3_projector} and \eqref{H3_compact_project} together with the ansatz for the gauge fields in \eqref{S3_gauge_fields} and \eqref{H3_gauge_fields}, we obtain the BPS equations          
 \begin{eqnarray}
f'&=& -\frac{1}{6\Sigma} \left(2 g_2-6\kappa a e^{-2 g} \Sigma^2-\sqrt{2} g_1 \Sigma^3\right),\\
g'&=& -\frac{1}{6\Sigma} \left(2 g_2+6\kappa a e^{-2 g} \Sigma^2-\sqrt{2} g_1 \Sigma^3\right),\\
\Sigma'&=& -\frac{1}{3} \left(g_2-3\kappa a e^{-2 g} \Sigma^2+\sqrt{2} g_1 \Sigma^3\right).
\end{eqnarray}         
These equations turn out to be the same as in the $SO(3,1)$ case after setting all the scalars from vector multiplets to zero. A single $AdS_2\times H^3$ critical point is again given by \eqref{AdS2_SO3_1}.                        
              
\section{Conclusions and discussions}\label{conclusion}
We have found a new class of supersymmetric black strings and black holes in asymptotically $AdS_5$ space within $N=4$ gauged supergravity in five dimensions coupled to five vector multiplets with gauge groups $U(1)\times SU(2)\times SU(2)$, $U(1)\times SO(3,1)$ and $U(1)\times SL(3,\mathbb{R})$. These generalize the previously known black string solutions preserving eight supercharges by including more general twists along $\Sigma_2$. Furthermore, unlike the half-supersymmetric solutions which only exhibit hyperbolic horizons, the $\frac{1}{4}$-supersymmetric black strings can have both $S^2$ and $H^2$ horizons. On the other hand, the $AdS_5$ black holes only feature $H^3$ horizons.
\\
\indent For $U(1)\times SU(2)\times SU(2)$ gauge group, we have identified a number of $AdS_3\times \Sigma_2$ solutions preserving four supercharges. The solutions have $U(1)\times U(1)\times U(1)$ and $U(1)\times U(1)_{\textrm{diag}}$ symmetries and correspond to $N=(0,2)$ SCFTs in two dimensions. We have given many examples of numerical RG flow solutions from the two supersymmetric $AdS_5$ vacua to these $AdS_3\times \Sigma_2$ geometries. We have also found a supersymmetric $AdS_2\times H^3$ solution describing the near horizon geometry of a supersymmetric black hole in $AdS_5$. For $U(1)\times SO(3,1)$ and $U(1)\times SL(3,\mathbb{R})$ gauge groups, all $AdS_3\times \Sigma_2$ and $AdS_2\times H^3$ solutions exist only for vanishing scalar fields from vector multiplets and have the same form for both gauge groups.  
\\
\indent It would be interesting to compute twisted partition functions and twisted indices in the dual $N=2$ SCFTs compactified on $\Sigma_2$ and $\Sigma_3$. These should provide a microscopic description for the entropy of the aforementioned black strings and black holes in $AdS_5$ space. On the other hand, it is also interesting to find supersymmetric rotating $AdS_5$ black holes similar to the solutions found in minimal and maximal gauged supergravities \cite{5D_rotating_BH1,5D_rotating_BH2} or black holes with horizons in the form of a squashed three-sphere \cite{5D_BH_Squashed_1,5D_BH_Squashed_2,5D_BH_Squashed_3}. Furthermore, embedding these solutions in string/M-theory is of particular interest and should give a full holograpic interpretation for the RG flows across dimensions identified here.         
\vspace{0.5cm}\\
{\large{\textbf{Acknowledgement}}} \\
P. K. is supported by The Thailand Research Fund (TRF) under grant RSA5980037.

\end{document}